# RW Aur A : SpeX Spectral Evidence for Differentiated Planetesimal Formation, Migration and Destruction in an ~3 Myr Old Excited CTTS System

C.M. Lisse[1], M.L. Sitko[2], S.J. Wolk[3], H.M. Günther[4], S. Brittain[5], J.D. Green[6], J. Steckloff[7,8], B. Johnson[9,10], C.C. Espaillat[11], M. Koutoulaki[12], S.Y. Moorman[13], A.P. Jackson[14]

[1]Planetary Exploration Group, Space Department, Johns Hopkins University Applied Physics Laboratory, 11100 Johns Hopkins Rd, Laurel, MD 20723   carey.lisse@jhuapl.edu

[2]Space Science Institute, Boulder, CO 80301, USA  sitko@SpaceScience.org

[3]*Chandra* X-ray Center, Harvard-Smithsonian Center for Astrophysics, 60 Garden Street, Cambridge, MA, 02138  swolk@cfa.harvard.edu

[4]Massachusetts Institute of Technology, Kavli Institute for Astrophysics and Space Research, 77 Massachusetts Avenue, NE83-569, Cambridge, MA 02139   hgunther@mit.edu

[5]Department of Physics & Astronomy, 118 Kinard Laboratory, Clemson, SC 29634  sbritt@clemson.edu

[6]Space Telescope Science Institute, 3700 San Martin Dr. Baltimore, MD 21218 jgreen@stsci.edu

[7]Planetary Science Institute, Tucson, AZ 85719  jordan@psi.edu

[8]Department of Aerospace Engineering and Engineering Mechanics, University of Texas at Austin, Austin, TX, 78712-1221 steckloff@utexas.edu

[9]Department of Earth, Atmospheric, and Planetary Sciences, Purdue University, West Lafayette, IN 47907

[10]Department of Physics and Astronomy, Purdue University, West Lafayette, IN 47907  bcjohnson@purdue.edu

[11]Department of Astronomy, Boston University, Boston, MA 02215  cce@bu.edu

[12]Dublin Institute for Advanced Studies, 31 Fitzwilliam Place, Dublin 2, Ireland  mariakout@cp.dias.ie

[13]Department of Physics, University of Texas at Arlington, Box 19059, Arlington, TX 76019 sarah.moorman@mavs.uta.edu

[14]School of Earth and Space Exploration, Arizona State University, 781 E Terrace Mall, Tempe, AZ 85287 alan.jackson@asu.edu

37 Pages, 16 Figures, 2 Tables

# Abstract

We present 2007 – 2020 SpeX VISNIR spectral monitoring of the highly variable RW Aur A CTTS. We find direct evidence for a highly excited, IR bright, asymmetric, and time variable system. Comparison of the spectral and temporal trends found determines 5 different components: (1) a stable continuum from 0.7 - 1.3 μm, with color temperature ~4000K, produced by the CTTS photospheric surface; (2) variable hydrogen emission lines emitted from hot excited hydrogen in the CTTS's protostellar atmosphere/accretion envelope; (3) hot CO gas in the CTTS's protostellar atmosphere/accretion envelope; (4) highly variable 1.8-5.0 μm thermal continuum emission with color temperature ranging from 1130 to 1650K, due to a surrounding accretion disk that is spatially variable and has an inner wall at r ~ 0.04 AU and T~1650K, and outer edges at ~1200K; and (5) transient, bifurcated signatures of abundant Fe II + associated SI, SiI, and SrI in the system's jet structures. The bifuracted signatures first appeared in 2015, but these collapsed and disappeared into a small single peak protostellar atmosphere feature by late 2020. The temporal evolution of RW Aur A's spectral signatures is consistent with a dynamically excited CTTS system forming differentiated Vesta-sized planetesimals in an asymmetric accretion disk and migrating them inward to be destructively accreted. By contrast, nearby, coeval binary companion RW Aur B evinces only (1) a stable WTTS photospheric continuum from 0.7 - 1.3 μm + (3) cold CO gas in absorption + (4) stable 1.8-5.0 μm thermal disk continuum emission with color temperature ~1650K.

**1. Introduction.** The main phases of early stellar growth occur when a nascent star is still deeply embedded inside its parent molecular cloud. The "Class II YSO" T Tauri phase of protoplanetary disk evolution and stellar growth occurs after the majority of the cloud's mass has condensed into a central object of ~ 1 $M_{solar}$ and the rest of its mass has collapsed into a surrounding dense disk of mass ~0.01 - 0.1 $M_{solar}$. Material from this disk actively accretes onto the star's surface. This phase will typically last < 2.5 Myr (Mamajek 2009 and references therein), and accretion variability is expected during these later stages of protostar evolution. This epoch is now also thought to be, from age-dating and dynamical studies of our solar system, the time period in which large differentiated planetesimals and giant planets first formed (Kruijer *et al.* 2014, Yang & Johansen 2014, McClure 2019). Therefore the study of T Tauri star evolution, like that incurred by our own proto-Sun, *is also* the study of planetesimal and planet formation in a solar system, and connecting up the two often disparate lines of inquiry (T Tauri stellar behavior and planetesimal formation, growth, differentiation, and migration) is important to get a complete physical picture of early solar system formation and evolution. In this paper we describe how we took a well studied T Tauri stellar system, RW Aur A, and by obtaining 0.7 – 5.0 µm spectra typically used to study solar system objects and planet hosting stars over the timespan 2006 - 2020, have found strong evidence for a stable protostar hosting a dynamically excited accretion disk currently forming, migrating, and disrupting Vesta-sized differentiated planetesimals.

**2. RW Aur A System Background.** RW Aur A (HD 240764) is a classical T Tauri star (CTTS), i.e. a well-formed protostar surrounded by a massive accretion disk, located at 160 pc distance (Bailor-Jones *et al.* 2018). The system is thought to contain an ~1.4 $M_{solar}$ protostar with luminosity ~ 1.6 $L_{solar}$ and age ~3.5 My (Dodin *et al.* 2020 and references therein) and a reported K3 spectral type. The object exhibits strong optical spectral signatures due to protostar photospheric emission, XUV emission due to accretion, and thermal emission from a surrounding thick accretion disk (Section 3). It has extended associated outflow jets, making RW Aur A one of the very few T Tauri stars with a known, resolved jet

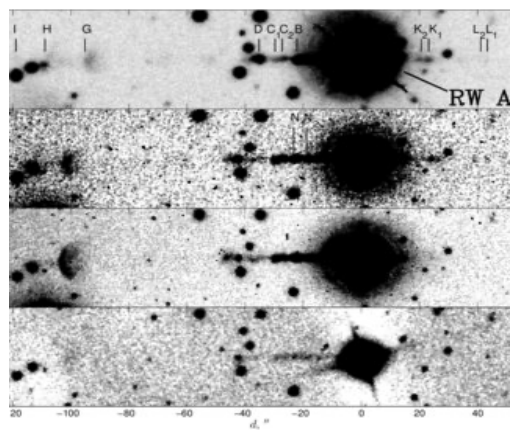

**Figure 1** – The jets of RW Aur A, imaged in S II (Mund & Eisloffwel 1998), and in S II, H-alpha, and Fe II by Berdnikov *et al.* (2017).

(Dougados *et al.* 2000; López-Martín *et al.* 2003; Beck *et al.* 2008; Fig. 1). It hosts a dense circumstellar disk of radius ~60 au (Cabrit *et al.* 2006; Rodriguez *et al.* 2018), about as large as the solar system's Kuiper Belt. RW Aur A is known to be a heavily accreting system, with dM/dt ~ $10^{-8}$ to $10^{-7}$ M$_\odot$/yr (Valenti *et al.* 1993, Hartigan *et al.* 1995, White & Ghez 2001, Alencar *et al.* 2005, Facchini *et al.* 2016, Gárate *et al.* 2019, Koutoulaki *et al.* 2019; for scale, in 1 Myr a 5 x $10^{-8}$ M$_\odot$/yr RW Aur A will accrete a further 0.05 M$_\odot$ onto the central protostar, ~4% of its final mass, or about 40x the mass of $M_{solarsytem\_planets}$ = 0.0013 M$_\odot$). Once RW Aur A finally settles down onto the main sequence, it is thus likely to be an early-F star of mass ~1.7 M$_\odot$.

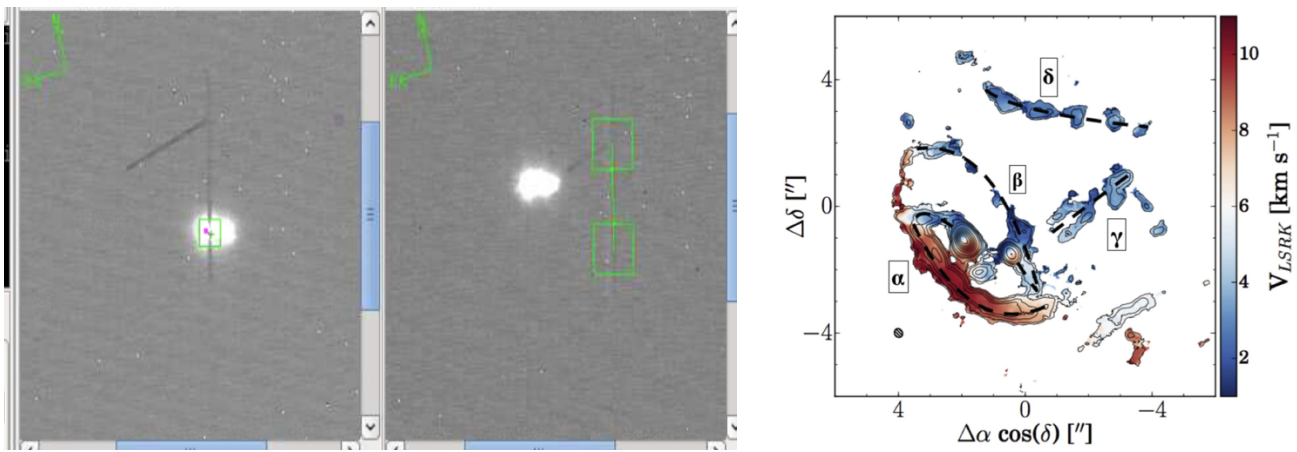

**Figure 2 – The RW Aur AB field.** *(Left)* RW Aur A (Eastern brighter object) and B (fainter object to the West) are located very close to each other on the sky according to the NASA/IRTF SpeX Guidedog camera, but can be easily separated spectrally using the fine 0.3" SpeX slit (vertical thin black line). The green boxes denote the extent of the telescope ABBA nods used to remove sky emission. *(Right)* The ALMA high resolution $^{12}$CO image of the same field by Rodriguez *et al.* 2018 shows a wealth of structure due to unconsolidated gas and dust and close binary companion RW Aur B, including two "classical tidal arms" (α and β in the figure) tracing out the orbit of the B component [at approx. (0, -1.5)] around A [at approx. (2,-1)] and two "detached tidal arms" γ and δ. The classical tidal arms were reproduced by Dai *et al.* (2015) through modelling of the dynamical encounter of the two protostars.

RW Aur A is one of the few T Tauri stars that exhibits the CO bandhead in emission due to hot gas in the innermost regions of the disk (Carr 1989; Beck *et al.* 2008; Eisner *et al.* 2014, Koutoulaki *et al.* 2019). RW Aur A is also a member of a close binary (~1.5" or 240 AU semi-major axis, Berdinkov *et al.* 2017) with weak-line TTS (WTTS), RW Aur B of reported mass ~0.85 M$_{solar}$ (Dodin *et al.* 2020 and references therein; Fig. 2). The RW Aur AB binary itself is located in the crowded young Taurus star forming region.

RW Aur A is well known (since the mid-1800's) to be a highly variable system with frequent multi-magnitude drops in apparent brightness lasting months to years (Chou et al. 2013, Dodin *et* *al.* 2019; Fig 3a). Its most recent dramatic dimming events suddenly started about 2010 (Rodriguez *et al.* 2013, 2016; Petrov *et al.* 2015; Fig. 3b). During the first modern dimming event, RW Aur A's V-band brightness dropped by 1.5-2 mag over almost 180 days (Rodriguez *et al.* 2013), but then recovered. The second modern dimming event, starting in 2014, was even larger, at more than two optical magnitudes (Petrov *et al.* 2015; Bozhinova *et al.* 2016), and is still affecting the system (Fig. 3b). The dimming events of RW Aur are similar (1-3 mag in depth) to those of UX Ori stars (Grinin *et al.* 1991), which are typically associated with dramatic accretion events in earlier type objects. A few other T Tauri stars undergo similar dimming events, for instance, AA Tau (Bouvier *et al.* 2013) and V409 Tau (Rodriguez *et al.* 2015).

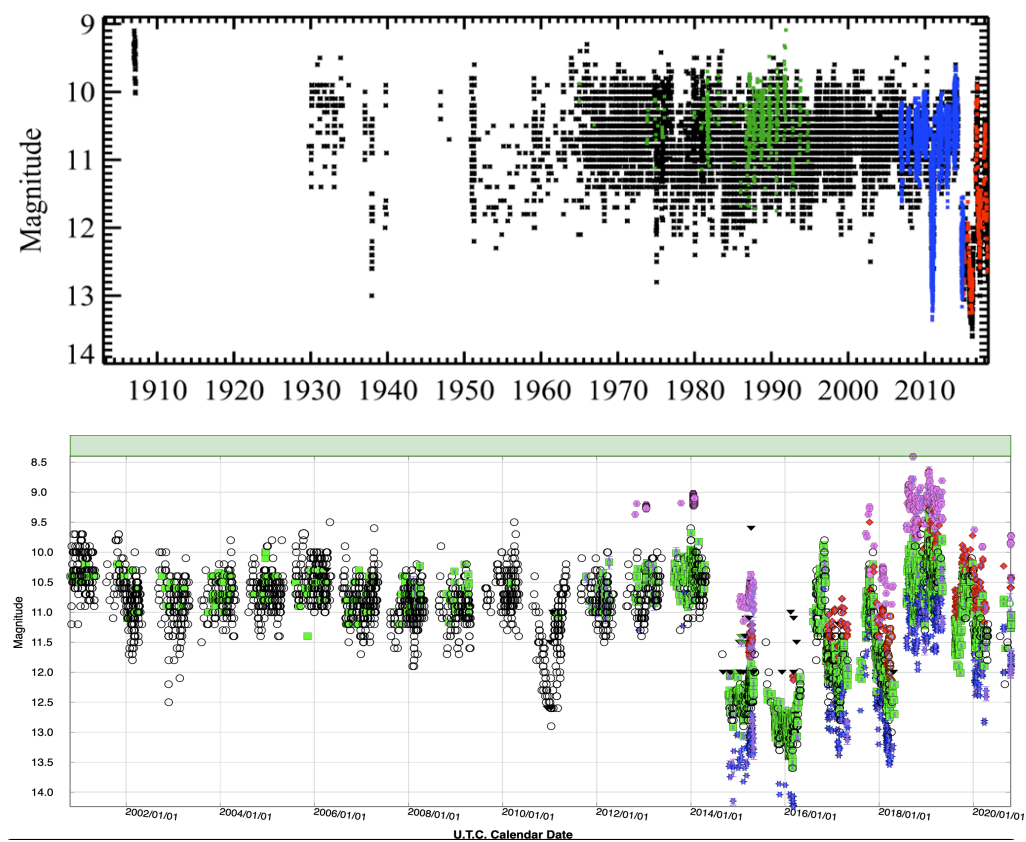

**Figure 3 – (*Top*) Optical lightcurve monitoring of RW Aur A** from 1906 to 2016, after Rodriguez *et al.* 2013, showing AAVSO (black), Wesleyan (green), KELT (blue), and ASA-SN (red) optical photometry trending for the system. (***Bottom***) Detailed optical lightcurve of RW Aur A from AAVSO from 2001 – 2020, showing relatively stable behavior (around a ± 0.5 mag measurement scatter) from 2001 – 2010, a strong, short dip in 2010, more stable behavior in 2011 – 2013, and then the onset of the recent steep drop from 2014 – 2019. The system may be returning to "normal" in 2020.

The origin of these dimming events is still debated. Popular hypotheses include inner disc warps, puffed-up inner rims, and starspot driven changes in the protostar's luminosity. RW Aur is a unique system that may shed light on the origin of this phenomenon, because of the large amount of data that is available before and during the dimming events, with both photometric and spectroscopic monitoring. For RW Aur A the occulting material has been associated with the inner regions of the disc due to additional emission (excess) at near-IR wavelengths and more particularly in L and M bands (Shenavrin *et al.* 2015; Bozhinova *et al.* 2016; Fig. 4).

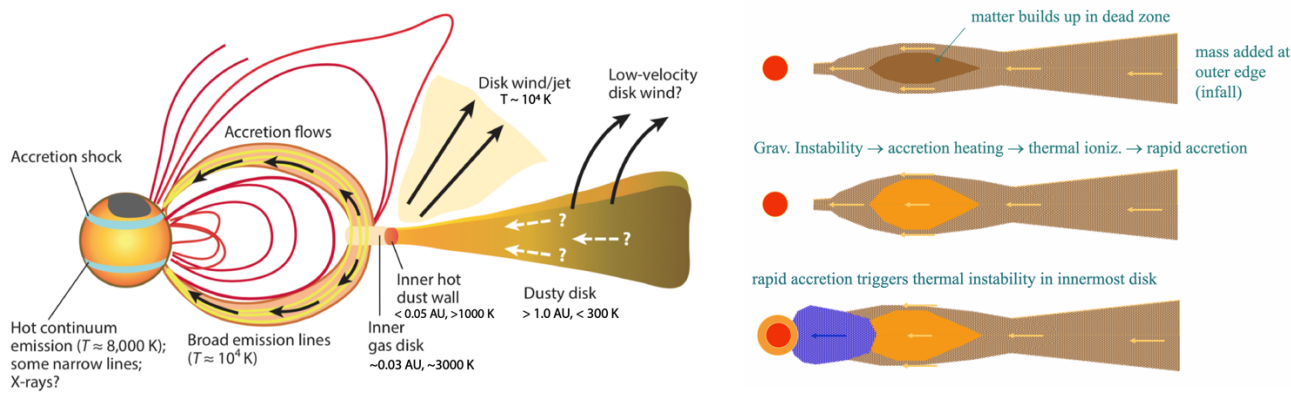

**Figure 4 – (*Left*) Best current understanding of CTTS log-scale structure, illustrating where different SED flux emissions are sourced.** The stellar photosphere has a similar 0.3 - 1.3 um spectrum to main sequence stars, with the addition of significant XUV from regions of shocked accretion deposition. The CCTS atmosphere/corona is comingled with the inward accretion flow of hot material and sources VISIR emission lines. The hot inner wall typically dominates near-IR continuum emission from the accretion disk at 1.5 - 4.0 um , but this can be obscured if large, optically thick amounts of outer, cooler disk material in an asymmetric disk move into the line of site. Bipolar jet outflow is also sourced near the inner disk wall. Outer accretion disk regions at distances beyond ~1 AU dominate the mid- to far-infrared, > 5 um spectral signature of the object. For a sense of scale, note that in this work the inner disk wall and co-rotation radii are at ~0.04 AU respectively, or 8-9 $R_\odot$, at the top of the modern day Sun's corona, while the accretion disk's outer edges extend out to ~100 AU. After Hartmann *et al.* (2016). (*Right*) Schematic of what happens during an outburst/lightcurve drop of the system, showing the overloading and breakdown of the "typical" magnetic field funneled accretion flow and filling of the innermost system regions. After Armitage (2001, 2002).

In 2018, a Chandra group led by Günther *et al.* (2018) announced the disappearance of the soft X-ray emission typically seen from RW Aur A and other T Tauri stars (Güdel *et al.* 2007, Skinner & Güdel 2014), and its replacement by strong emission by iron (hereafter Fe) at 3-6 keV (Figure 5). Further analysis by this group concluded that the region around the CTTS had been filled with large amounts (~$10^{18}$-$10^{19}$ kg) of Fe material[1], most likely by the destruction of a planetesimal core (Gárate *et al.* 2019). Soft x-ray absorption and harder x-ray emission in the presence of optical extinction has been seen before, in strongly accreting T Tauri stars and Fu-Ori objects (Skinner *et al.* 2006, 2010; Liebhart *et al.* 2014). For example, the soft source in DG Tau has been identified with an X-ray jet close to the star, while Güdel *et al.* (2007) attribute the hard source to a corona with much higher gas column densities than expected from visual extinction and interstellar gas-to-dust mass ratios. The models proposed to explain these observations either involve dust-depleted accretion streams from the disk to the star or dust-depleted winds launched from the inner disk.

---

[1] Using mass attenuation coefficients values for Fe of μ = 9.1e3 cm$^2$/g at 1 keV, 1.6e3 at 2 keV, 5.6e2 at 3 keV, 2.6e2 at 4 keV, 1.4e2 at 5 keV, and 8.5e1 at 6 keV (https://physics.nist.gov/PhysRefData/XrayMassCoef/ElemTab/z26.html), I = I$_o$e$^{-(\mu*column)}$, and assuming that the Fe fills a spherical region the size of the inner CO gas disk 0.03 AU in radius or a sphere inside the co-rotation radius of 0.042 AU, one needs ~ 2 mg/cm$^2$ total Fe column (or 3 to 6 x $10^{18}$ kg of Fe) in order to attenuate the 1-2 keV X-rays by a factor of 50 or more, while also allowing 3 to 6 keV X-rays to still emerge.

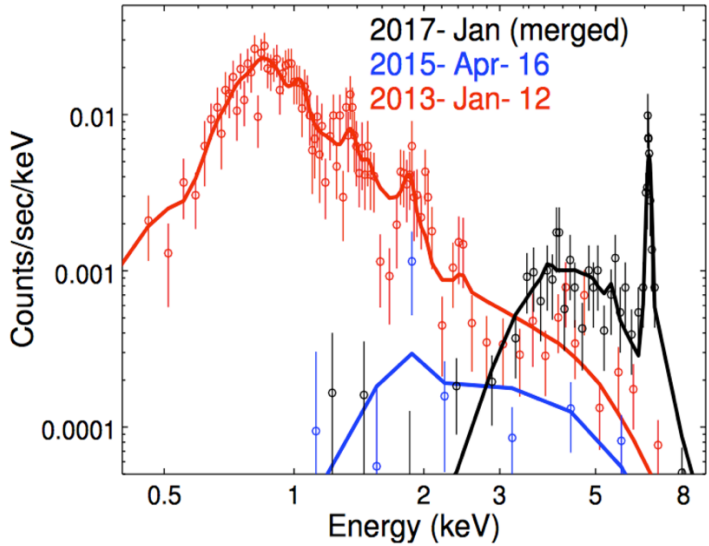

**Figure 5 – Chandra monitoring results of the RW Aur A system**, showing the marked change from "normal" soft X-ray emission seen on 12 Jan 2013, likely sourced by the accretion shock regions on the protostellar photospheric surface (**red**), to the enshrouded and highly extincted x-ray emission seen on 15=6 Apr 2015 (**blue**), and then progressing to a new regime of hard X-ray, Fe-dominated emission seen in Jan 2017 (**black,** merged coaddition of multiple visits). Some of the same Fe must have been delivered to the proto-star's corona in order to produce the new strong K-alpha Fe emission peak seen at ~6.3 keV (after Wilms *et al.* 2000 and Günther *et al.* 2018).

Whether RW Aur A's new Fe was produced in 2015 via a planetesimal-planetesimal collision in the system's accretion disk or via accretion + vaporization of a planetesimal core in the regions just adjacent to proto-stellar surface's accretion spots was not clear from their measurements. Comparing X-ray observations in RW Aur A's bright and dim states, Schneider *et al.* (2015) found an increase in the gas column density during the second dimming event. Additional X-ray observations by Günther *et al.* (2018) also showed an increased gas column density as well as an iron enhancement in the stellar corona. They studied the extinction from optical (Antipin *et al.* 2015) to NIR (Schneider *et al.* 2015), and found that it is nearly grey, suggesting grain growth in the inner disc. Other authors (see e.g. Petrov *et al.* 2015) suggested that the extinction could be due to an outburst or a wind. Facchini *et al.* (2016), on the other hand, argued that these dimming events are due to a perturbation of a misaligned or warped inner disc.

For all of these observed characteristics, RW Aur A appears to be an outstanding laboratory for understanding the physical mechanisms related to late stage protostar growth and protostellar disk evolution. Knowing that there are important FeI and FeII absorption/emission lines in the near-infrared, an ongoing 0.8-5.0 μm NASA-IRTF 3.3m/SpeX spectral characterization study of exoplanet and exodisk host stars (the NIRDS study; Lisse *et al.* 2015, 2017a,b; 2020) was thus repurposed to measure the emission lines from RW Aur A. Having obtained SpeX spectra of 90+ stars with $2 < K < 12$ at ~1% relative flux precision down to declinations of -50 degrees from 2010 to 2020, this was readily done for bright RW Aur A (K ~ 7) at +30 degrees declination using the IRTF located at 19 degrees north latitude. Further, co-author and chief observer M. Sitko of the NIRDS study had used the same SpeX instrument to monitor RW Aur A in 2006 and 2007, providing a long temporal baseline when combined with the new measurements. Here we present the results of these new measurements, compare them to other reported measurements in the literature, and a provide a migrating differentiated planetesimal synthesis model consistent with all the reported measurements.

**Table 1. Program Observations of the RW Aur A CTTS System**

| Telescope/Instrument | λ (μm) | Observation Date (UTC) | Cal Star | Mode |
|---|---|---|---|---|
| ASAS/AAVSO/ Kelt/Wesleyan | 0.55 | 01 Jan 2001 – 20 Oct 2020 | | CCD photometry |
| IRTF/SpeX | 0.8 – 5.5 | 30 Nov 2006 | HD 31295 | SXD + LXD |
| | 0.8 – 2.4 | 25 Feb 2007 | HD 31295 | SXD |
| | 0.7 – 5.0 | 29 Nov 2018 | HD 34203 | SXD + LXD |
| | | 23 Mar 2019 | HD 34203 | SXD + LXD |
| | | 29 Sept 2019 | HD 34203 | SXD + LXD |
| | | 06 Mar 2020 | HD 34203 | SXD + LXD |
| | | 07 Oct 2020 | HD 34203 | SXD + LXD |
| VLT/XSHOOTER | 0.3 – 2.46 | 19 Mar 2015 | HD 80781 | UVB/VIS/NIR |
| | | 30 Sept 2016 | HD 20001 | |

**3. Observations.** We first observed RW Aur A on 13 Jan 2006 UT from the NASA/IRTF 3 m telescope on the summit of Mauna Kea, HI using SpeX. The "old" SpeX instrument provided R = 1800 to 2200 observations from 0.7 – 5.2 μm in two orders, termed SXD (for "short cross-dispersed") and LXD (for "long cross-dispersed") when configured with the narrowest 0.3" slit (Rayner *et al.* 2003; 2009). Post-2012, this instrument was updated with a new focal plane and slit optics to "new" SpeX configuration, providing roughly the same R = 2000 to 2500 observations from 0.7 to 5.0 μm.

The observational setup was identical to that used by Sitko in his early work pre-main sequence star work (e.g. Sitko *et al.* 2012) and for our more recent NIRDS debris disk & planet hosting stellar spectroscopy studies (e.g. Lisse *et al.* 2015, 2017a,b, 2020), although care had to be taken to use the narrowest slit possible to exclude the nearby (1.2" distant) RW Aur B binary companion by centering the slit on RW Aur A with alignment perpendicular to the RW Aur A – B axis (Fig. 2). The nearby A0V star used in Spextool data reduction as a calibration standard (Vacca *et al.* 2004) was HD 31295 (K=4.4) in 2006-2007, and HD 34203 ($K$ = 5.46) in 2018 – 2020. Both stars were picked to match RW Aur A ($K$ = 7.4) in brightness and angular proximity on the sky. RW Aur A was typically observed in SXD mode with a total on-target integration time of 960 sec and in LXD mode for 1800 sec, while the calibrator star was observed for 480 sec in SXD and 1440 sec in LXD mode. Both target and calibrator were observed using ABBA nod patterns to remove telescope and sky backgrounds. At +30 degrees declination, RW Aur was easily observed at low airmass from Mauna Kea at +19 degrees latitude. We typically observed RW Aur A in early February – mid-March within a few hours of morning twilight, and 7 months later in late September – early October it was observed a few hours after the end of evening twilight. The observational dataset consists of 7 measurements of RW Aur A taken over the time period 2006 - 2020 (Table 1). During this time, the object has gone from a quiescent, stable state (lasting at least from 2001 thru 2013 according to optical photometric monitoring; Fig. 3) through a period of semi-chaotic lightcurve dips in 2014 - 2019, and back to the beginnings of normalcy in 2019-2020.

Also included in the analysis, in order to help fill in the large gap in time between the first 2006/2007 SpeX observations and the modern era 2018 – 2020 observations, are archival XSHOOTER 0.29 – 2.46 μm observations reported by Koutoulaki *et al.* 2019 (Fig. 6). These observations have R = 5000 to 8000 in the NIR, a few times better than SpeX, and extend shortwards down to 0.3 μm in the near UV but only extend out to 2.46 μm in the near IR. The 2015 XSHOOTER measurements look to have comparable SNR to the SpeX observations and better spectral resolution, but the 2016 XSHOOTER data demonstrates multiple satellite lines unseen in any of the other observations and is likely plagued by optical ghosts, so we use XSHOOTER data here only to determine the overall shape of the bright continuum.

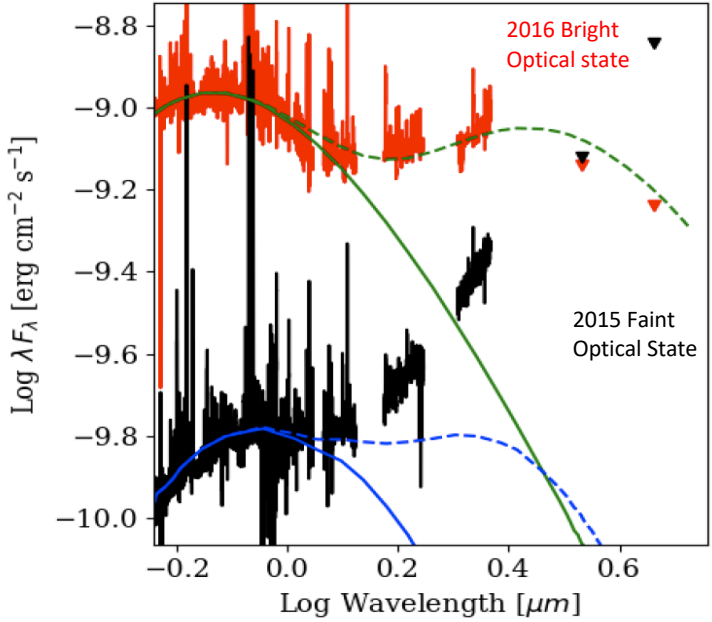

**Figure 6 – 0.60 to 2.46 μm SED of RW Aur A, as measured by Koutoulaki *et al.* 2019 using VLT/XSHOOTER.** The stable short wavelength, 4900K stellar photosphere contribution and the highly variable long wavelength, accretion disk thermal emission contributions (**black**, 2015 faint optical lightcurve state; **red**, 2016 bright lightcurve state) are readily apparent. Triangles – WISE 3.6/4.5 μm photometry for the system in the bright state (red) and faint state (black). Green solid curves – 4900K "veiled" blackbody fit to the photospheric continuum. Green dashed line – 4900K photosphere + 1180K reddened disk continuum model contribution. Blue solid, dashed curves - Reddened 4900K blackbody fit to the photospheric continuum. Green dashed line – photosphere + 1180K reddened disk continuum model contribution. The anti-correlation of the optical photospheric and NIR disk continuum is readily apparent. The veiled 4900K photosphere + 1180 K accretion disk model is consistent with the ~4000K unveiled + 1280 K (faint state) or 1650 K (bright state) accretion disk fits we find from the SpeX observations (Fig. 7).

We made one other group of measurements of importance to this study. Episodically, we observed the binary companion TTS RW Aur B, located within 1.2" of prime target RW Aur A (and thus very close by in the 1' SpeX guider field of view and easily found, Fig. 2; all that was required was to relocate the 0.3" x 15" slit by ~5 slit-widths). The RW Aur B observations showed a stable object evincing flux from a photosphere, a moderately bright surrounding accretion disk, but very little atomic hydrogen line emission (Fig. 7). It has a weak accretion disk thermal signature, which does not obscure the central protostar, and all its characteristic non-hydrogen atomic and molecular lines are in absorption, indicating relatively cold source populations. RW Aur B looks much more like a "typically behaved" ~3 Myr old, very young WTTS (i.e., a growing planet-hosting system with most of its protostellar accretion completed) than RW Aur A. These results are consistent with the compiled findings of Dodin *et al.* 2020, who report B as a low accretion rate UX-Ori-like object with $T_{eff}$ = 4100−4200 K, $A_V$ = 0.6±0.1, $L_*$ ≈ 0.6 $L_\odot$, $R_*$ ≈ 1.5 $R_\odot$, M ≈ 0.85 $M_\odot$, $\dot{M}_{acc}$ < 5×10$^{−9}$ $M_\odot$ yr$^{−1}$.

Since the RW Aur A and B components are closely bound and likely coeval, this directly demonstrates how asymmetric the AB system is. It also demonstrates how driven and excited the A component must be to express such strong atomic and hot molecular v' = 2 CO emission lines (and thus appears to be an EXOri-like object with excited disk surface gas; Connelley & Reipurth 2018) and evince such a highly variable, bright accretion disk (Dai *et al.* 2015; Fig. 8).

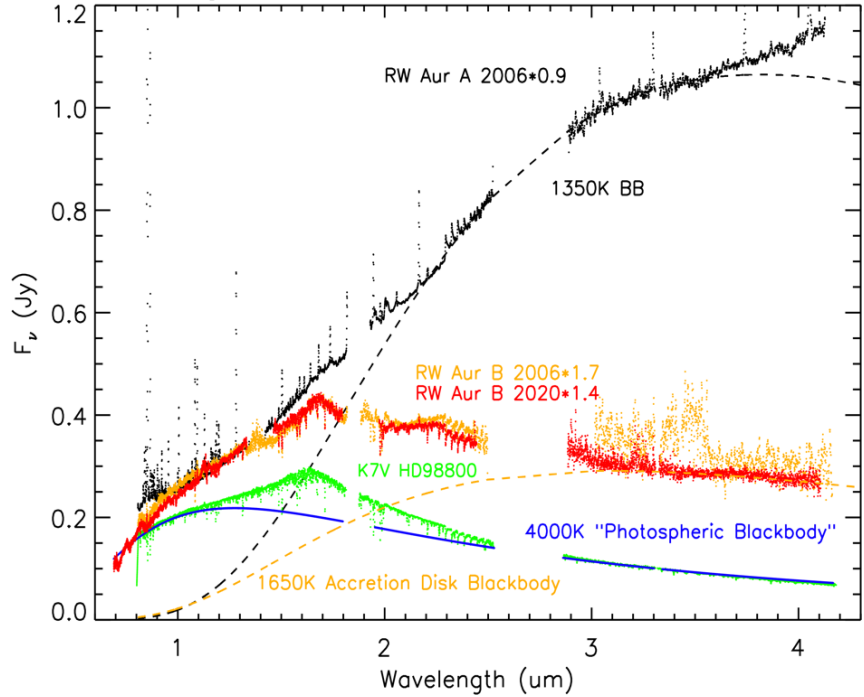

**Figure 7** – 2006 SpeX RW Aur A spectrum (**black**) versus the 2006 SpeX spectrum for coeval binary companion RW Aur B (gold), taken during the early bright, relatively quiescent phase of RW Aur A's behavior using "old" SpeX. Also shown (**orange**, **red**) is a 2020 spectrum of RW Aur B taken with the improved "new" SpeX instrument, demonstrating the overall stability of its spectral structure. Both spectra show roughly 4000K stellar photospheric emission patterns at the shortest 0.7 – 1.0 µm wavelengths. The "hump" at ~1.65 µm in the B spectra is a typical of cooler star spectra, as demonstrated by the overplotted HD98800 K7V spectrum (**green**) from the NIRDS survey. RW Aur A evinces multiple pronounced atomic emission lines, seen as the strong narrow spikes above the continuum from 0.8 to 3.3 µm; RW Aur B does not. RW Aur A also shows the ro-vibrational spectral emission pattern of hot CO gas from 2.2 to 2.4 µm (Section 4.3), while RW Aur B shows the CO ro-vibrational complex at 2.2 – 2.4 µm in absorption due to the presence of cold CO gas. The accretion disk thermal emission for RW Aur A is cold, strong, and significant down to ~1.4 µm, while the thermal emission from B's disk is hot but faint and important mainly out past 1.8 µm. In all ways the RW Aur A system seems much more energetically excited than its coeval RW Aur B binary partner.

**4. Results.** Here we present the results of our 0.7 – 5.0 µm optical/near-infrared spectral measurements. The measurements are most sensitive to hot material of temperature > 1000 K residing within 1 au of the RW Aur A protostar, and thus probe the innermost portions of the system's accretion disk, its protostellar atmosphere and surface, and its outflow jets (Fig 4).

**4.1 NIR Continuum + Atomic Emission Lines.** In Figures 8 through 11 we show the results of IRTF time domain spectral monitoring of RW Aur A from 2006 - 2020. The large grasp spectrum of Fig 8 shows how the continuum emission can evolve over time: the shortest wavelength portion, from 0.8 through 1.5 µm, identified with the protostar's photospheric emission, stays relatively constant in shape (although its amplitude can change, indicating that whatever additional material is added to increase the overall extinction, it is grey and thus likely consists of > 10 µm dust particles). By contrast, the longer wavelength continua, identified with thermal emission from the inner wall of the accretion disk (as previously seen in other TTS,

Muzerolle *et al.* 2003, Espaillat *et al.* 2010, McClure *et al.* 2013), is highly spectrally variable and shows multiple changes from 2006 through 2020.

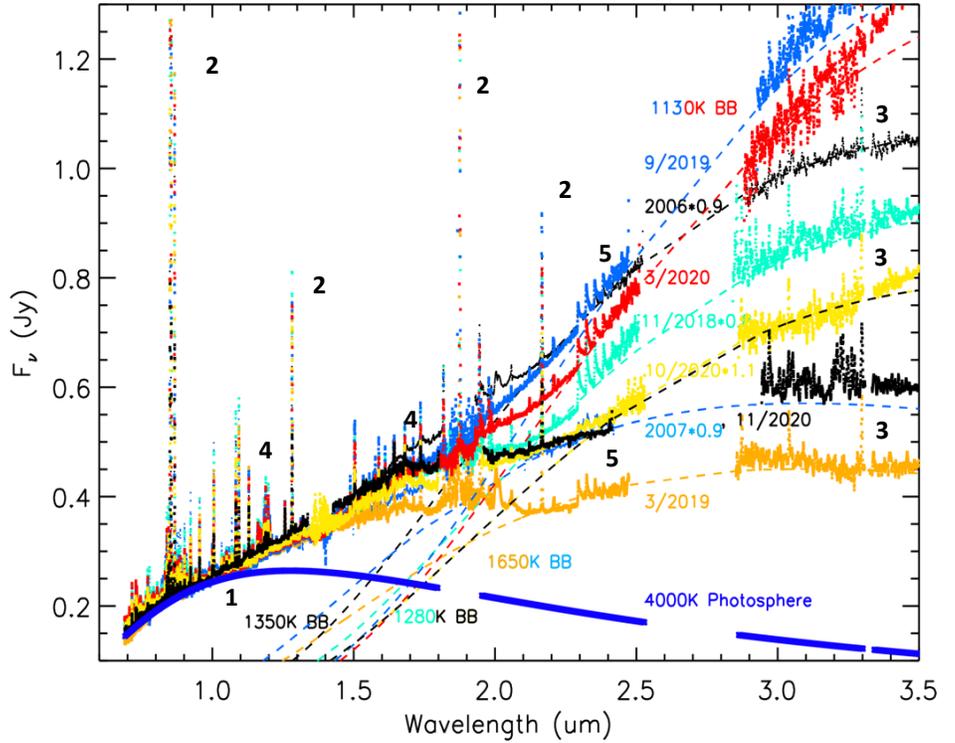

**Figure 8 – Rich 0.8 to 3.5 μm SpeX spectra of RW Aur A taken from 2006 through 2020** (Table 1). Empirically, five flux components are seen in these spectra: (1) a short wavelength continuum from 0.7 – 1.3 μm with approximately constant spectral shape over time; (2) very strong singly H-Paschen lines with variable amplitude; (3) a long wavelength component that varies between a low flux, high color temperature state and a high flux, low temperature state; (4) weak emission lines of Fe II and associated species (see text) that change between single peaked, near-zero amplitude and doubly peaked, large amplitude states; and (5) strong ro-vibrational line complex emission from hot CO gas. The "baseline, quiescent" state of (3) is characterized by the 2007, 3/2019, and 10/2020 spectra: a moderately low level of 1.8 – 3.3 μm flux coupled with a relatively high color temperature ~1650K plus a bright state for components (1) and (2). The "puffed up accretion disk" state, as seen in 2006, 9/2019, and 3/2020 demonstrates high levels of cold (~1300K) emission (3), low levels of H-Paschen emission (2), and lower optical flux levels (1). The quiescent vs. puffed-up temporal trends for components (1,2,3) are very different than that found for the weak bifurcated line emissions (4).

These accretion disk emission excursions have structure, however: the quiescent state appears to be characterized by high color temperature but low total flux, while the stochastic/excited state demonstrates low color temperature but high total flux. (Our current understanding of this is that the accretion disk (Fig. 4) occasionally "puffs up" dramatically so that the outer, colder parts of the disk obscure the much hotter inner disk edge from view (Section 5).) The observed color temperatures of the disk thermal emission, ranging from 1130 K (highly perturbed) to 1650 K (quiescent), argue for material at 0.04 to 0.1 AU from the protostar's center (assuming the protostars' current luminosity is 1.6 $L_{solar}$). The first dramatic increase of lower color temperature disk emission in our dataset was seen in 2006, long before the Fe X-ray onset of 2015 that triggered this study, suggesting that the accretion disk variability and new Fe creation are not coupled. At the same time, we noticed important temporal patterns in the shape of the system's emission lines.

The Hydrogen Paschen lines in the monitoring data were all found to be single peaked, with amplitudes that varied inversely with the long wavelength continuum flux (Figs. 9 – 10). The CaII triplet lines were also found to follow this trend.

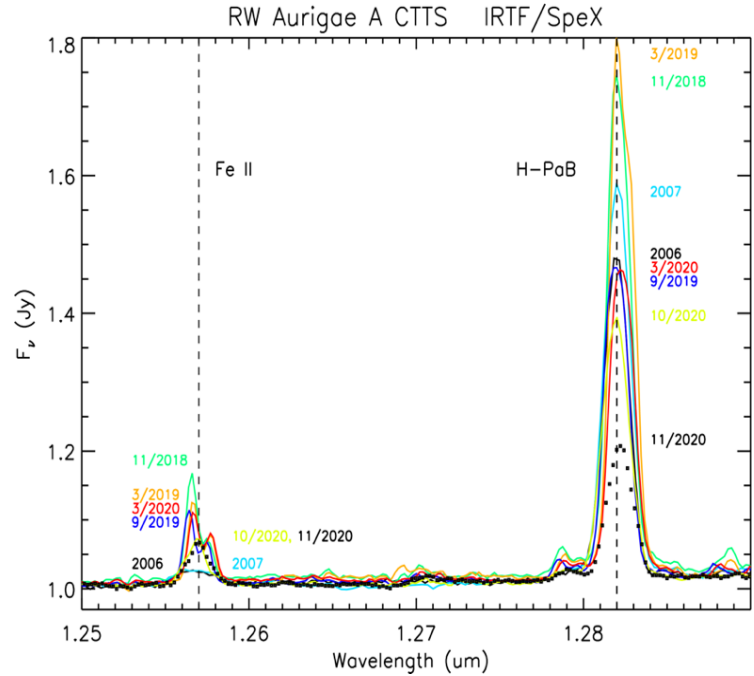

**Figure 9 – IRTF/SpeX 1.257 μm Fe II line monitoring results for RW Aur A**, showing the marked change in the Fe II content of the system from near-zero, single peaked, stable state of 2006/2007 behavior to the strong, highly bifurcated Fe-rich state of Nov 2018 – 3/2020. The bifurcation splitting is large, denoting red/blue shifts of 100-200 km/sec. By 10/2020, however, the bifurcation structure had disappeared, although the single peaked line centered on 1.257 μm is still much greater in amplitude than seen in the 2006/2007 stable lightcurve state data. By contrast, the nearby single-peaked H-Paschen β emission line, while varying significantly in amplitude throughout the time range of observation, does not decrease monotonically in amplitude from 2018 through 2020. Instead the H-Paschen β emission amplitude anti-correlates with the long-wavelength 2.2 – 3.5 μm accretion disk flux (Fig. 8), suggesting that the H-Paschen β becomes obscured when the colder outer regions of the disk puff up and obscure emission from the hottest innermost regions.

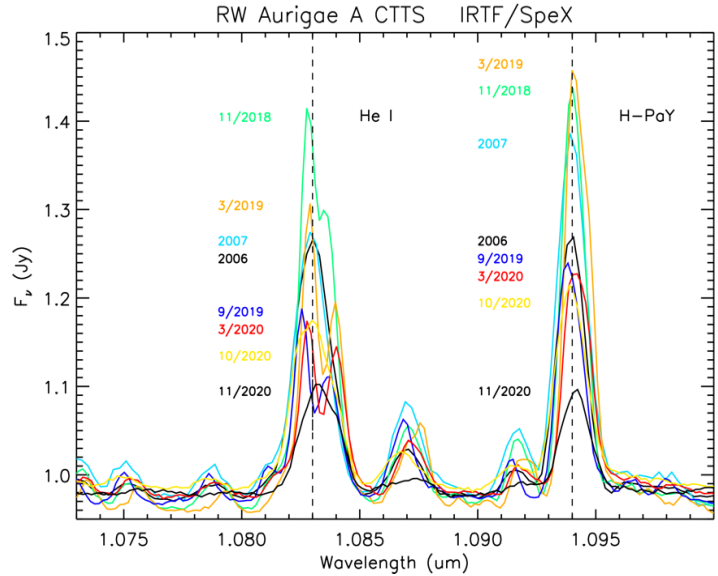

**Figure 10 – IRTF/SpeX 1.083 μm HeI line monitoring results for RW Aur A,** showing the grow-in of bifurcated emission line structure in the 2018 – 2019 timeframe, as for Fe II. Unlike Fe II, however, there was a sizable single-peaked emission line component for HeI in 2006/2007, arguing that the HeI emission in 11/2018 – 3/2020 was sourced from two different regions of the system. By contrast, the H-Paschen γ line at 1.094 μm, while varying significantly in amplitude throughout the time range of observation shows only single peaked behavior throughout. The H-Paschen γ emission amplitude anti-correlates with the long-wavelength 2.2 – 3.5 μm accretion disk flux of Fig. 8, again suggesting that the H-Paschen γ becomes obscured when the colder outer regions of the disk puff up and obscure emission from the hot inner regions.

By contrast, the strong new Fe II lines found in the 2018 – 2020 data (and also in the 2015 – 2016 XSHOOTER data of Koutoulaki *et al.* 2019) have a very different temporal signature: they are nonexistent in 2006/2007, strong and bifurcated from 2015 through 2019, decreasing in amplitude monotonically with time and finally disappearing to leave a single peaked feature by late 2020 (Fig. 9).

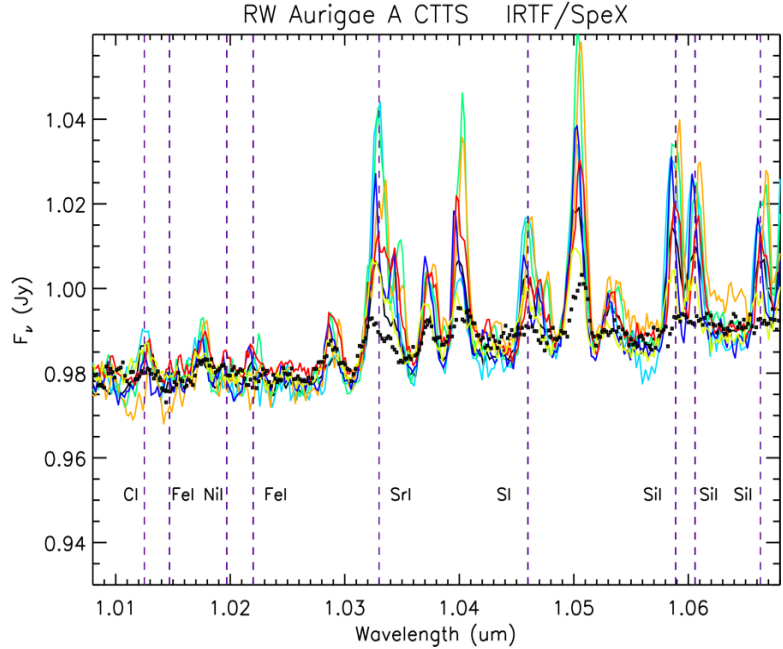

**Figure 11 – IRTF/SpeX 1.040 – 1.063 μm monitoring results for RW Aur A,** showing the temporal behavior for a spectral region typically rich in unblended stellar absorption/emission lines. SiI at 1.029 μm, SrI at 1.033 μm, and SI at 1.046 μm show the same kind of behavior as HeI 1.083 μm. By contrast, FeI 1.0149 μm, NiI at 1.0197 μm, and FeI at 1.022 μm do not. The nearby Paschen-Delta line at 1.005 μm shows the typical, single peaked behavior of the other Paschen hydrogen lines. This argues for SiI, SrI, and SiI being present in the material supplying the additional Fe II of 2018 – 2020.

We used the FeII lineshape temporal trends to search for other emission lines in the 2006 - 2020 spectral dataset that track Fe. The same temporal signature was found for bifurcated emission lines of HeI (Fig. 10) and SiI, SrI, and SI (Fig. 11). The usually present emission lines of AlI, FeI, MgI, or NaI were not detected. This argues for SiI, SrI, and SiI being present in the material supplying the additional Fe II of 2018 – 2020. Si and S are expected at the few percent level in the makeup of a planetary core (Section 5.2), consistent with the Günther *et al.* 2018 model for sourcing of the abundant Fe via destruction of a differentiated planetesimal core. Carbon is as well, and while we do have a positive detection of CI at 1.025 μm in this Fig. 11, the SNR of the detection is too low to determine if its temporal signature matches the others. Like AlI, MgI, and NaI, SrI is a lithophilic element not expected in a core, so its presence implies some rocky material present as well – although we must allow for the fact that it is a very efficiently emitting species, like the chemically very similar Ca.

**4.2 Fe II Line Analysis.** A least 5 good lines due to Fe II emission at 0.862, 1.257, 1.321, 1.534, and 1.644 μm were seen in the SpeX spectra (Fig. 12). All were bifurcated, and are indicative of hot (T = 10,000 to 20,000 K), shocked Fe II ions. These lines are the brightest ones utilized by Koo *et al.* (2016, Figs. 2-4), and we follow their Fe II emission line analysis, which uses Fe II line ratios, widths, and amplitudes, coupled with their shocked plasma radiative models, to determine plasma parameters. The resulting RW Aur A jet plasma diagnostics determined from the measured FeII line ratios, widths, and amplitudes (Fig. 12; Table 2) indicate a very dense and hot emitting Fe II environment ($n > 10^6/cm^3$, T = 15,000 to 20,000 K, $v_{relative}$ ~ 200 km/sec). (Note that the LSRK velocity of RW Aur A, ~6 km/sec [Rodriguez *et al.* 2018; Fig.2b], is small compared to the jet velocity and we ignore it here.) This temperature range and relative velocity vs. the local ISM are consistent with that found by Berdnikov *et al.* (2017) for RW Aur A's jets, but our new density estimate is almost two orders of magnitude larger. Either the emission is coming from a highly localized and densified region (i.e., an outflowing impulsive outburst), or there is something wrong with our analysis. The presence of an outflowing Fe II emitting clump, though, with the proper emission time circa 2015 has been imaged by Takami *et al.* 2020, making a short term, rapidly moving jet density enhancement plausible. Following Crovisier (2002), a lower Fe II ion number and mass estimate for the potentially optically thick emitting Fe II in the clump can be derived using the Einstein A coefficients quoted in Koo *et al.* 2016, assuming the Fe II emission is being driven by a 20,000K blackbody shock heat bath surrounding each Fe II ion.

To start, the excitation rate g of a fundamental vibrational band at wavelength l by a blackbody of solid angle $\Omega$ and temperature $T_{ex}$ is given by:

$$g = [\Omega /4\pi] A_{21}/ [ exp( hc/\lambda kT_{ex}) -1]$$

where $A_{21}$ is the Einstein A coefficient of for the given Fe II line and h, c, and k are Planck's constant, the speed of light, and Boltzmann's constant. Assuming $[\Omega /4\pi] = 1$ for an Fe II ion fluorescing in a shocked jet clot due to photons in the surrounding jet heat bath at temperature $T_{ex}$, this reduces to

$$g = A_{Fe\ II\ line}/(e^{[hc/\lambda kT_{ex}]} -1)$$

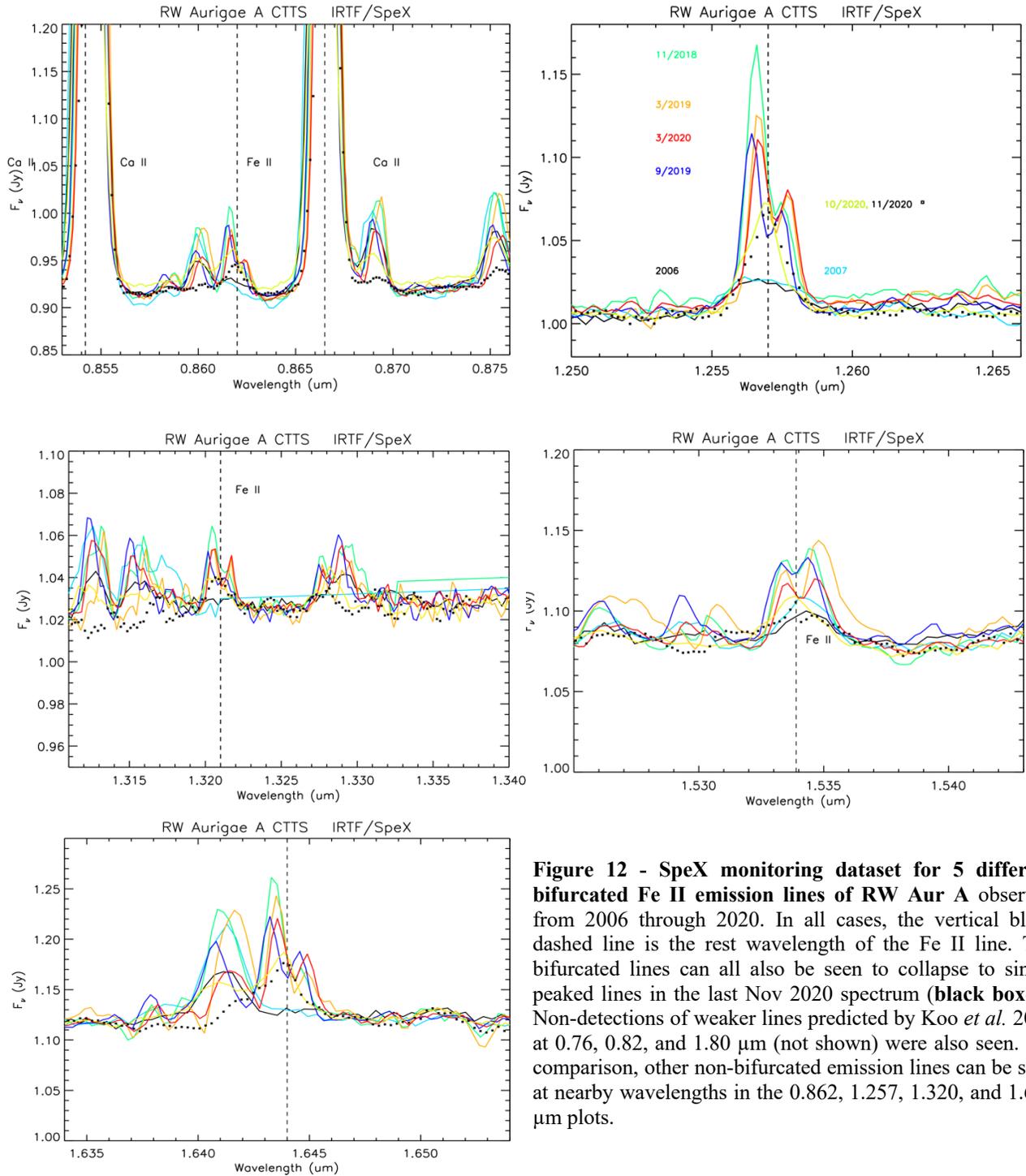

**Figure 12 - SpeX monitoring dataset for 5 different bifurcated Fe II emission lines of RW Aur A** observed from 2006 through 2020. In all cases, the vertical black dashed line is the rest wavelength of the Fe II line. The bifurcated lines can all also be seen to collapse to single peaked lines in the last Nov 2020 spectrum (**black boxes**). Non-detections of weaker lines predicted by Koo *et al.* 2016 at 0.76, 0.82, and 1.80 μm (not shown) were also seen. For comparison, other non-bifurcated emission lines can be seen at nearby wavelengths in the 0.862, 1.257, 1.320, and 1.644 μm plots.

For an optically thin jet (for NIR absorption coefficients of ~$10^{-3}$, this occurs for total Fe gas column densities $n_{Fe} \lesssim 10^{20}$ cm$^{-2}$), the number of Fe II ions can be calculated in a straightforward manor from the value of the g-factor and the observed strength of the IR line:

$$N_{Fe\,II} = 4\pi \Delta^2 F_\nu\, d\nu / (g * h\nu) = 4\pi \Delta^2 F_\nu\, d\lambda / (g * h\lambda)$$
$$= 4\pi * (160\ pc * 3.1 \times 10^{16} m\ pc^{-1})^2 * (1.0\text{e-}26\ Wm^{-1}Hz^{-1}Jy^{-1}) * F_\nu\ (Jy) * d\lambda\ (e^{[hc/\lambda kT_{ex}]} - 1) / (A_{Fe\,II\ line} * h\lambda)$$
$$= 3.1\text{e}12 * 0.1\ Jy * 0.002\ \mu m\ (e^{[1.44\text{e}4\ \mu mK/(1.257\ \mu m * 20000K)]} - 1) / (5.27\text{e-}3 * 6.626\text{e-}34 * 1.257)$$
$$= 1.2\text{e}44\ \text{ions at}\ 1.257\ \mu m\ \text{for}\ T_{ex} = 20{,}000\ K\ \text{and max line amplitude of Fig. 9}$$

where F is the flux in the line, $d\lambda$ is the width of the line, $\lambda$ is the line center wavelength of the band, g is the g-factor for the band, $N_{mol}$ is the number of molecules in the beam, and $\Delta$ is the Earth – RW Aur A distance. Converting from the number of Fe II ions to the mass of Fe II gas, assuming 56 amu per Fe II ion gas, we find for the 1.257 μm line

$$M_{Fe\,II} = N_{mol\,Fe\,II} * 56\ amu * 1.66 \times 10^{-27}\ kg\ amu^{-1}$$
$$= 1.2\text{e}44 * 56 * 1.66\text{e-}27$$
$$= 1.1 \pm 0.4\ (1\sigma) \times 10^{19}\ kg\ Fe\,II$$

Note that for varying decreasing $T_{ex}$ from 20,000 down to 15,000 K will only increase these numbers by ~30%, and that for an optically thick Fe II cloud, these estimates are lower limits to the true amount of Fe II in the emitting cloud.

The results of the lower limit mass determinations for each of the 5 Fe II lines are listed in Table 2. These should be compared to the minimum amount of Fe gas mass required to obscure the low energy X-rays normally detected by Chandra, $10^{18}$ - $10^{19}$ kg (Fig. 5; Günther *et al.* 2018). Taken at face value, a direct comparison of these two lower limits indicates that roughly equal amounts of the Fe newly created in 2014 was blown out of the system's jets and accreted onto the protostar, and the amount of FeII mass in the event was about that found in the large asteroids 16 Psyche or 4 Vesta.

**Table 2 : Lower Limit Mass Determinations for 5 Bifurcated [FeII] Line Detections[a]**

| Line Wavelength | Line Width $\Delta$(lambda) | Flux | Einstein $A_{21}$[b] | $T_{excitation}$ | Minimum [Fe II] Number | Minimum[c] Mass of [Fe II] |
|---|---|---|---|---|---|---|
| (μm) | (μm) | (Jy) | (sec$^{-1}$) | (Kelvin) | | (kg) |
| 0.862 | 0.0015 | 0.06 | 2.73e-2 | 20,000 | 0.25 e44 | 0.23 ± 0.08e+19 |
| 1.257 | 0.002 | 0.10 | 5.27e-3 | 20,000 | 1.2 e44 | 1.1 ± 0.4e+19 |
| 1.3209 | 0.0022 | 0.025 | 1.49e-3 | 20,000 | 1.0 e44 | 0.94 ± 0.4e+19 |
| 1.5339 | 0.0022 | 0.05 | 2.64e-3 | 20,000 | 0.81 e44 | 0.76 ± 0.3e+19 |
| 1.644 | 0.0025 | 0.07 | 5.07e-3 | 20,000 | 0.58 e44 | 0.54 ± 0.2e+19 |

[a]Total mass summed over both parts of the bifurcated line.
[b]Wavelengths, Einstein A-coefficients, and $T_{excitation}$ values from Koo *et al.* (2016)
[c]For comparison to the Fe II mass estimates, 16 Psyche is a core that masses 2 x $10^{19}$ kg, 4 Vesta masses 2.6x$10^{20}$ kg with an ~5x$10^{19}$ kg core, the Moon masses 3x$10^{22}$ kg with an ~3x$10^{20}$ kg core, Mercury masses 3.3x$10^{23}$ kg with an ~2.4x$10^{23}$ kg core, and the Earth , and the Earth masses 6 x $10^{24}$ kg with an ~2 x $10^{24}$ kg core.

**4.3 CO Molecular Emission.** With a low spectral resolution instrument like SpeX at R ~ 2000 it is nearly impossible to see the individual CO ro-vibrational lines of EX Ori-like (Kospal *et al.* 2011, Connelley & Reipurth 2018) RW Aur A in the fundamental around 4.7 µm and in the first overtone at 2.2 to 2.4 µm. But what can be seen at low resolution are the bandheads – and strong bandhead detections are good enough to make some strong physical conclusions. Fundamental ν' = 1 ro-vibrational CO emission from the inner disk of CTTS's is very common. There is no unbiased survey of CTTSs, but the detection rate is ~100% for v=1-0 emission lines near 4.7 µm, demonstrating that abundant CO is common in T Tauri disks like RW Aur B (Fig 6b; e.g., Najita *et al.* 2003; Salyk *et al.* 2011; Brown *et al.* 2013; Banzatti *et al.* 2018). On the other hand, populating the ν' = 2 transitions, like the ones we find for RW Aur A using SpeX, is very challenging (Figs. 7 & 8). The vibrationally excited bands can be populated through gas fluorescence by 1500 Å photons (e.g., Krotkov, Scoville, & Wang 1980; Brittain *et al.* 2007), but T ~ 4000 K CTTS photospheres do not produce these in abundance. More likely, these lines can be populated by collisions. The latter requires a critical density of $n(H) \sim 10^{14}$ cm$^{-3}$ and temperatures of ~2000 K in optically thin gas (e.g., Scoville, Krotkov, & Wang 1980; Najita *et al.* 1996). Additionally, the Einstein A coefficients of the ν = 2-0 transitions are two orders of magnitude smaller than the Einstein A coefficients for the ν = 2-1 transitions. Thus, CO emission from the first overtone (as opposed to the fundamental) of T Tauri stars is quite rare. Indeed, the first overtone CO bandheads are only observed in emission from disks in 23% of Class I YSOs (Connelly & Greene 2010; see also Carr 1989 and Chandler *et al.* 1993), and they are even rarer among Class II YSO's (e.g., Greene & Meyer 1995).

The fundamental CO spectrum of RW Aur A observed in January 2001 was presented in Najita *et al.* (2003). These authors found that the emission arose from a narrow annulus extending from 0.014 au to 0.14 au. The temperature and surface density of the CO are described with the power-laws $T(r_h) = 3500K (R/0.014)^{-0.3}$ and $N(CO) = 3 \times 10^{18}$ cm$^{-2}$ $(r_h/0.014)^{-0.3}$. These authors also note that the fit parameters for the overtone emission and fundamental emission are consistent.

To characterize the variability the CO emission arising from the inner disk, we fit previously unpublished high-resolution data acquired with Keck2/NIRSPEC on 22 March 2002 (Fig 13a). Following Najita *et al.* (2003), we fit the overtone spectrum (Fig. 13b) and find that the CO extends from 0.027±0.01 au to 0.18±0.02 au. The temperature and column density of the CO are described with the power laws $T(R)=3500^{+1000}_{-500}$ K $(R/0.03)^{-0.3}$ and $N(CO)=4.6\pm0.2 \times 10^{19}$ cm$^{-2}$ $(R/0.027)^{-0.3}$. We also find that to fit the power-law spectrum we have to account for overtone CO absorption from the stellar photosphere. We adopt a stellar temperature of 4000K and $v\sin(i)=18.6$ km s$^{-1}$ (White & Ghez 2001) and find we need a veiling of $r_K = 21\pm2$. The inner CO gas disk radius determined by modeling compares well with the r = 0.035 AU found for an inner accretion disk wall at 1650 K in radiative equilibrium with a central host protostar of 1.6 L$_{Solar}$ luminosity, and a dynamically inferred co-rotation radius of 0.042 AU for an 1.4 M$_\odot$ protostar rotating with 2.64 day period (Section 5.1), being just inside these radii as expected using current models of CTTS's (Fig. 4).

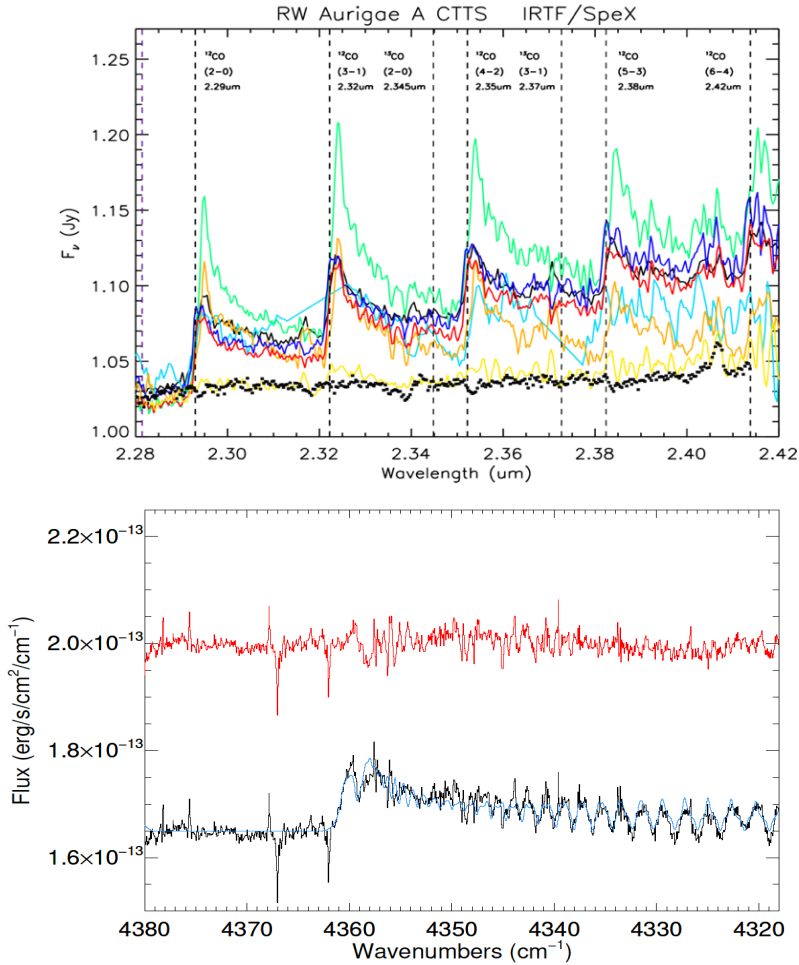

**Figure 13 – (*Top*) SpeX monitoring of RW Aur A's CO lines from 2006 to 2020.** The strength of the first overtone line complex emission is highly variable over time, being roughly stable from 2006 through 2019 (except for Nov 2018 (green), when it was twice as strong), but near zero by late 2020. (***Bottom***) **Best-fit hot gaseous CO 2-1 first overtone emission model (blue curve) versus high resolution Keck2/ NIRSPEC measurements (black) taken on 22 Mar 2002.** The red curve shows the (observations-model) residuals x 100. We have used R ~ 20,000 Keck2/NIRSPEC data for the CO analysis because the IRTF/SpeX R ~ 2000 data are not detailed enough to constrain the individual CO 2-1 emission peaks well. The inner CO gas disk radius determined by modeling compares well with the r = 0.035 AU we find for an inner accretion disk wall at 1650 K in radiative equilibrium with a central host protostar of 1.6 L$_{Solar}$ luminosity, and a dynamically inferred co-rotation radius of 0.042 AU for an 1.4 M$_\odot$ protostar rotating with 2.64 day period.

**4.4 Observing Results Summary.** To summarize the results of our 2006 - 2020 SpeX 0.8 - 5.0 μm observations of RW Aur A: by comparison to the spectral signature of close binary coeval companion RW Aur B, RW Aur A is a highly excited system at ~3 Myr, acting much more like a T Tauri star of a few hundred Kyr's age. Large amounts of new Fe II gas emission appeared by 2015, that had mostly disappeared by 2020. The single-peaked to bifurcated to single peaked temporal pattern of the Fe II line profiles (Figs. 9, 12, 14) allowed us to search for emission from other volatilized atomic species, like HeI (Fig. 10) and we also detected transient SI, SiI, and SrI (but not Al, Ca, Na, K, or Mg) coincident with the Fe II (Fig. 11). The strongly bifurcated appearance of the emission lines implies that the emitting gas is moving at +/- 100-200 km/sec with respect to the line of sight (LOS), suggesting that the emitting gas was in the bipolar outflow jets, a determination bolstered by the disappearance of the Fe II signal after ~6 yrs of flowing out of the SpeX input slit aperture.

We also see a strongly time variable accretion disk flux signature, providing direct evidence for an asymmetric or time variable outer accretion disk – but the temporal signature of the disk's variability does not correlate well with the Fe II line's variability (Fig. 14), nor any of the other three major flux sources (Fig. 8) producing RW Aur A's time variable spectrum over the 2006 – 2020 time period (Fig. 14).

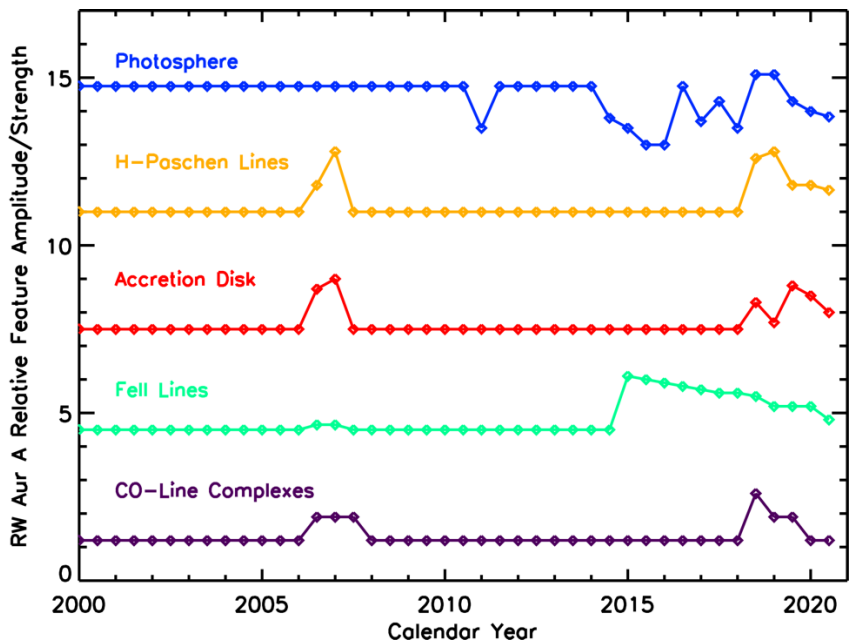

**Figure 14 – Summary plot showing the relative strength of the different flux sources of RW Aur A on a 6-month cadence similar to that we have employed for the IRTF monitoring observations.** All 5 sources show a different temporal signature, suggesting the fluxes are produced by different sources (e.g., from top to bottom, the stellar atmosphere, accretion disk, outflow jets, and the stellar photosphere). The optical dimming seen in lightcurves (Fig. 3) follows the Photospheric trending. The closest similarity, if there is any, is between the H-Paschen line and the accretion disk NIR thermal continuum strength, indicating that puffing up the accretion disk may correlate with increased excitement of the CTTS's atmosphere.

On the other hand, the obscuration of the normal soft X-ray signal by optically thick gas seen by Chandra from 2017 through (at least) 2020 suggests that at least some of the emitting material is close into the central protostar. The SpeX observations from 2018 through Fall 2020 also show a single-peaked Fe II component that is elevated over the 2006/2007 pre-Fe II state, so it is likely that hot Fe gas was emplaced into both the system's widespread outflow jets and close-in shocked T Tauri atmosphere.

One other important point concerning the observed variability of RW Aur A's flux components – they all vary on half-yearly to yearly timescales. This implies that they all have source functions variable on these timescales, and residence times for their reservoirs that are also on the order of years. Given that the dynamical parameters of the inner system are set by the Keplerian velocities in the inner accretion disk and the protostar co-rotation period to be on the order of a few days at speeds of 100's of km/s (Section 5.2), this is reasonable as 100's of orbits and rotations and motions over 10's of AU's are occurring in 1 years' time.

**5. Discussion.** In this section we present the physical aspects of the RW Aur A system most relevant to the new spectral measurements, deduce the implications of their temporal behavior, and then synthesize a physical model explaining both the aspects and their timing utilizing the current understanding of T Tauris stars and solar system formation.

**5.1 Basic Physical Picture of the System.** To understand the implications of the SpeX NIR spectral monitoring we first need to set the basic physical picture of the system, as much as it is known. RW Aur A is grouped as a classical T Tauri star (CTTS), meaning that it has contains about a solar mass worth of material, most of which is contained in a central small, condensed object but a reasonable fraction, a few tenths of a solar mass, of which is still contained in a thick circum-protostellar disk (Hartmann *et al.* 2016). Accretion of material onto the protostar occurs via funneling of disk material from the inner edge of the disk through magnetic conduits and onto hot spots on the protostellar surface (Fig. 4). The envelope of accretive material onto a CTTS is dense, likely optically thick in places, and rotating rapidly at its inner edge, which is locked via magnetic fields into co-rotating with the protostar. As RW Aur A has demonstrated a short optical lightcurve periodicity of 2.64 days (Petrov *et al.* 2001, Rodriguez *et al.* 2013) with a period consistent with the observed rotation periods seen for young FGK stars, this is thought to be the protostar's rotation period. Material moving with orbital period 2.64 days has the same orbital period as material in circular Keplerian orbits with orbital velocity of 173 km/sec at the co-rotation radius of 0.042 AU[2] around the 1.4 M$_\odot$ RW Aur A protostar. This co-rotation radius value is also very close to the 0.037 AU radiative equilibrium distance we find for an inner disk wall heated to T$_{LTE}$ = 1650 K by absorption of radiation from a 1.6 M$_\odot$ protostar. Lightcurve variability on the order of days to weeks (Fig. 14) can easily be created by the rotation of the central protostar itself, and an accretion disk with asymmetric large scale structure.

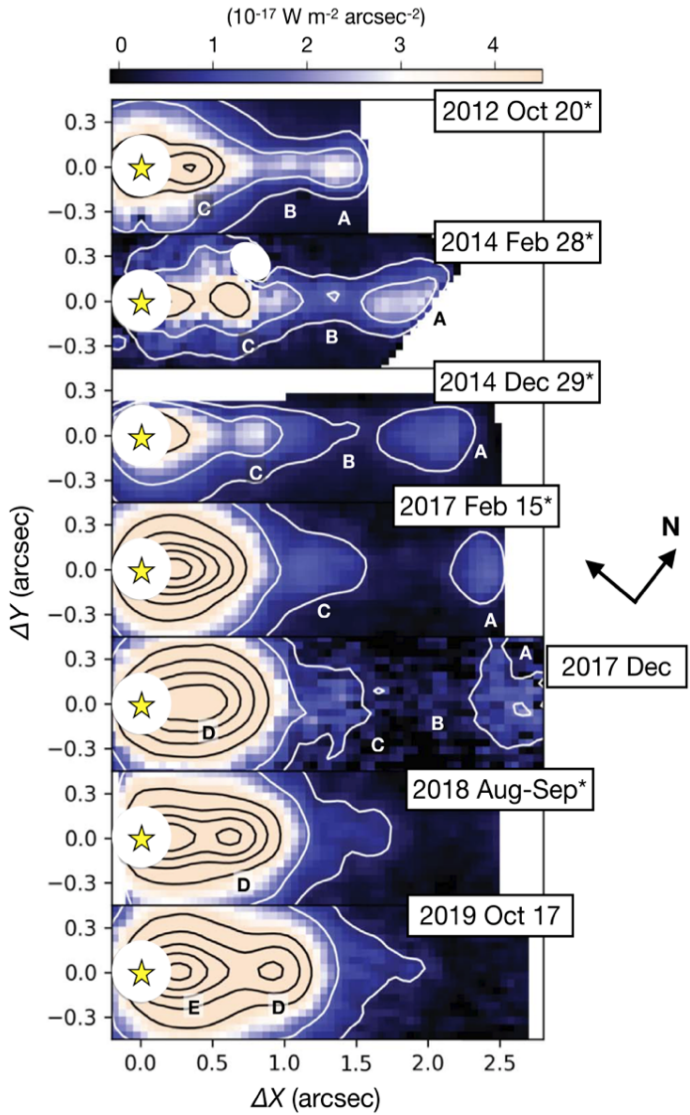

**Figure 15 – NIR imaging of the redshifted jet of RW Aur A in the 1.644 μm Fe II line from 2012 to 2019.** A large increase in the Fe II emission in the base of the jet is seen after 2014, coincident with the reported rise in hard Fe X-rays by Günther *et al.* (2018). Much of this brightness is seen to be flowing down the jet between 2017 and 2019, consistent with our SpeX bifurcated FeII line observations. Figure after Takami *et al.* (2020).

Accretion from the disk onto the protostar happens stochastically - while there does seem to be a rough steady state amount of material being dumped daily onto the condensed objects surface, large variations from steady state levels are often seen. The accretion disk likely contains not only a relatively smooth collection of micron-sized dust particles and gaseous species, but also streaming instability densifications and condensed planetesimals as well (Kruijer *et al.* 2014, Yang & Johansen 2014, McClure 2019) - some of which occasionally spiral their way down to the inner edge of the accretion disk and interact with dense collections of material at the accretion funnels before accreting onto the surface of the protostar.

---

[2] For context, 1 R$_\odot$ = 0.0047 AU, and the modern day Sun's corona extends out to ~10 R$_\odot$, so RW Aur A's inner disk wall and co-rotation radius would be located inside it, about 8 times closer than our planet Mercury is to the Sun.

The accretion funnels are not fully understood, but seem to also be associated with large outflow jets. Conservation of energy and angular momentum demands that in order to accrete material onto the protostellar surface, other material has to be ejected, carrying away the excess angular momentum and energy. Two strong "exhaust jets" have been known for decades in RW Aur A, having been imaged out to 1000's of AU from the protostar's center in hot gas atoms like Fe II (Melnikov *et al.* 2009, Takami *et al.* 2020; Figs. 1, 15).

RW Aur A's marked bipolar jets are not seen in the average 3 Myr old TTS, nor are they seen in coeval WTTS RW Aur B. Strong bipolar jets are usually seen in YSO and protostar systems within the first 0.1 to 0.5 Myrs, before the accretion disk settles down into a stable steady state phase of convectively driven moderate accretion onto the central condensation and the rotating star's magnetic field and wind can carry away the necessary angular momentum (Andre & Montmerle 1994, Frank *et al.* 2014, Bally 2016, Pudritz & Ray 2019).

Following Dai *et al.* 2015, we hypothesize that the excess energy exhibited by the A system versus the B system is coming from the dynamical excitation that the B binary member + surrounding ISM material is imparting to A. That is, the energy emitted in the A system's atomic lines, in its accretion disks thermal radiation, in its hot CO gas emission, and in its X-rays, is being input from the kinetic energy and angular momentum of the wide binary orbit. Further, we note that the high rate of accretion seen for the system (~ $2 \times 10^{-8}$ to $1 \times 10^{-6}$ $M_\odot$/yr, (White & Ghez 2001; Valenti *et al.* 1993; Hartigan *et al.* 1995; Gárate *et al.* 2019; Koutoulaki *et al.* 2019) and its estimated age of ~3 Myr (Dodin *et al.* 2020) combine to suggest that, in addition to energy, material from the ISM is entering the system. Without this infalling material the ~0.2 $M_\odot$ accretion disk would be quickly depleted and the current highly variable behavior and high accretion rates of RW Aur A would be very ephemeral and short-lived. This could also explain why the A system looks like the B system with added colder, but dynamically excited, outer-disk material. Planetesimal formation in young solar systems is expected in the first few Myr, especially in systems with streaming instabilities and density enhancements created by traveling sound waves (Bitsch *et al.* 2015, Wahlberg & Johansen 2017 and references therein). In such a system, it is not surprising to see stirred planetesimal ***formation and migration*** from the outer to the inner regions as deduced from our observations. The ultimate fate of such a  system is likely to be a non-Laplacian, non-HL Tau like collection of planets and a host star massing significantly more than its coeval B partner.

**5.2 RW Aur A's Response to an Impulse.** In light of this physical picture for RW Aur A, we can suggest two possible scenarios that can explain the SpeX spectral trending behavior we have found (Fig. 14). Both scenarios have in common a CTTS protostar in the latter stages of its growth with luminosity driven by accretional deposition of mass from a few tenths of a solar mass circumstellar accretion disk with inner wall at ~0.04 AU (Fig. 4). The location of the light sources from the protostar's photosphere, the inner gas disk, and the surrounding accretion disk (spectral components 1 through 3 discussed above) are the same in both. Both create variability in the optical photospheric emission, hot hydrogen gas line emission, and accretion disk thermal emission as the disk puffs up and relaxes (or has varying amounts of cold outer disk density enhancements rotating through the LOS). What differs between them is where the transient Fe II/SiI/SrI/SI gas is thought to have been in the system from 2014 - 2020.

**5.2.1 Jet Outflow & Exhaustion.** In this scenario the Fe II/SiI/SrI/SI jet lines were created in an impulsive event in late 2014 via a massive collision near the inner edge of the accretion disk. This material was subsequently rapidly injected into the system's bipolar outflow jets. Jet material moving at 150 km/sec since late 2014 is expected to leave the FOV of SpeX's slit after ~6 years. The total apparent motion of jet material first introduced in 2014 should be, by late 2020, 2.6 x $10^{10}$ km or 176 AU. Allowing for the ~45° tilt of the jets to the LOS (Berdnikov *et al.* 2017, Bailor-Jones *et al.* 2018, Doudin *et al.* 2020), this results in a projected on the sky distance of 125 AU. Assuming a 160 pc distance to RW Aur A (Bailor-Jones *et al.* 2018), this is equivalent to 0.8" on the sky. 0.8" is very much outside the 0.3" slit width employed in the SpeX monitoring, but is roughly equivalent to the effective seeing of 0.8 – 1.1" conditions encountered in the observations (Table 1). Thus material introduced into the bipolar jets in a delta function spike in late 2014 should be clearing out of the SpeX slit FOV by late 2020. This scenario can easily explain the shape of the bifurcated spectral lines, and their eventual disappearance. It is also entirely consistent with the marked brightening in FeII imagery of the jets by Takami *et al.* (2020) from 2017 through 2019 and movement down the jet of a bright clot that has reached ~0.9" from the center of the system by late 2019 (Fig. 15). However, it does not easily explain the marked optical and soft x-ray flux decreases of 2015 – 2019: the ability of fast-moving jet material to obscure the soft X-ray signal and replace it with a hard Fe-signal in the Chandra monitoring (Fig. 5; Günther *et al.* 2018 and references therein) would only have lasted for weeks to months as the material left the innermost regions of the system.

**5.2.2 Continual Inner Wall Supply.** An alternative scenario that supposes that the Fe II was continually sourced at ~0.04 AU in the innermost system from late 2014 through mid-2020, but then quickly de-excited in jet shocks so as to be removed from the SpeX FOV. This scenario can easily explain the low level, single peaked Fe II emission seen in 2020, can explain the optical lightcurve drop and soft x-ray diminution. But it runs afoul of the Gemini/VLT observations by Takami *et al.* (2020) of a discrete excess "blob" of hot, excited material traveling down the system's bipolar jets out to 1000's of AU (Fig. 15). Further, in order to explain the slow monotonic decrease in bifurcated spectral line shape amplitude that we found from 2015 – 2020 (Figs. 9, 12, 14), the material put into the winds & jets cannot have remained in the inner system where it would be obscured by the accretion disk's puffing up, like the H-Paschen lines are.

**5.2.3 Synthesis: Inner System Gas Reservoir Enrichment + Jet Outflow & Turnoff.** A more unified model supposes that the Fe II + associated species were created in late 2014, and injected into the system's outflow jets (Fig. 1), ***as well as*** into the hot gas reservoir interior to the accretion disk above the photospheric surface (Fig. 4). This is consistent with the physical picture that in order to conserve energy and angular momentum, some material has to be accelerated and ejected at the inner wall as other material is deaccelerated and eventually accreted onto the protostar. It is also consistent with the findings of a number of studies showing that impulsive accretion and jet outflow events are coupled (Facchini *et al.* 2016, Koutoulaki *et al.* 2019).

The Fe II not incorporated in the jets but emplaced in the protostar's atmosphere must effectively cover and obscure the hot XUV emitting accretion spots on the protostar's surface from Earth view in order to produce the effects seen by Chandra (Günther *et al.* 2018). It is reassuring that the estimate in Section 2 of $10^{18}$ to $10^{19}$ kg of space-filling Fe material obscuring the protostar's low energy x-ray emission is very similar to the amount of new Fe II mass estimated to have appeared in the system's jets from late 2018 to early 2020 (Section 4.2). It is also likely that this obscuring protostellar atmospheric material is still there according to the later Nov 2020 SpeX measurement – the single peak now seen is about 1/3 the EW area of the bifurcated peak structure, but it is still much more prominent than in the 2006/2007 spectra, which provides a useful baseline. This peak now has the same broad +/- 200 km/sec width demonstrated by the single peaked H-Paschen lines, which we attribute to the range of velocities expected for a uniform spherical distribution of emitters co-rotating with the protostar. Since we know that the H-Paschen lines are modulated by the accretion disk "puffing up" (Section 4, Figs 8 – 10) they must be sourced from down deep inside the disk as well.

There are three important ways to test this unified scenario. The first way is to search for an independently varying single peaked residual in the 2015 – 2020 Fe II 1.257 μm line shape, and see if its amplitude correlates with the accretion disk's "breathing" (in fact a 3-peak Lorentzian fit to the bifurcated Fe II structures, using 2 very narrow Lorenztians + 1 broad H-Paschen width Lorentzian, shows that the broad feature could indeed have been there throughout the 2015 – 2020 highly variable period (Figs. 3, 8). The second method is to continue monitoring the system spectrally in the near-IR in conjunction with Chandra as we have been doing, and see if the small single Fe II 1.257 μm peak persists as long as the Chandra Fe X-rays are found, but finally disappears when the Chandra Fe X-ray opacity disappears. The third way is to continually spectrally monitor the system at high cadence over the course of one 2.64 day rotation of the CTTS.

**5.3  Migrating Planetesimal Sourcing of the New Fe.** In this section we present a simple physical "toy" model, guided by current understanding of planetesimal and stellar formation, that uses a migrating, Fe core containing (i.e., differentiated) planetesimal as the source of the new Fe detected by SpeX and Chandra in RW Aur A's atmosphere and jets. The model is consistent with the fact that Fe II is produced at $10^4 – 10^5$ during highly energetic events, like giant impacts or shock heating in the outflowing jets and stellar atmospheric accretion region of RW Aur A. It is also consistent with current solar system formation model expectation that migrating, differentiated planetesimals with sizes and masses ranging from Vesta to Mars condense and travel through protoplanetary disks in the first few Myr of a solar system's life (Chambers 2004, Kenyon & Bromley 2006, Asphaug & Reufer 2014, Minton & Levison 2014, Baruteau et al. 2016), and that the minor C and Si species found in our SpeX bifurcated line observations of the system are commonly found at the few % level vs Fe in planetary cores.  We then analyze the energetics, dynamics, implications, and caveats of the toy model

**5.3.1 Toy Model : Catastrophic Disruption of a Differentiated Planetesimal in the Inner Accretion Disk.** In this physical model, a differentiated Vesta-sized planetesimal, about the size of the smallest oligarchs in models of hierarchical planetary formation (c.f. Chambers 2004, Kenyon & Bromley 2006, Asphaug & Reufer 2014, Minton & Levison 2014, Baruteau et al. 2016) is catastrophically and completely pulverized via a giant impact in the innermost regions of the RW Aur A accretion disk. Differentiated body formation and subsequent migration via core-accretion or streaming instability is expected (Kruijer *et al.* 2014, Liu & Ji 2020) in a massive asymmetric disk as clearly stirred and spectrally excited as RW Aur A is by its dynamical orbital interaction with RW Aur B (Figs. 2, 7).

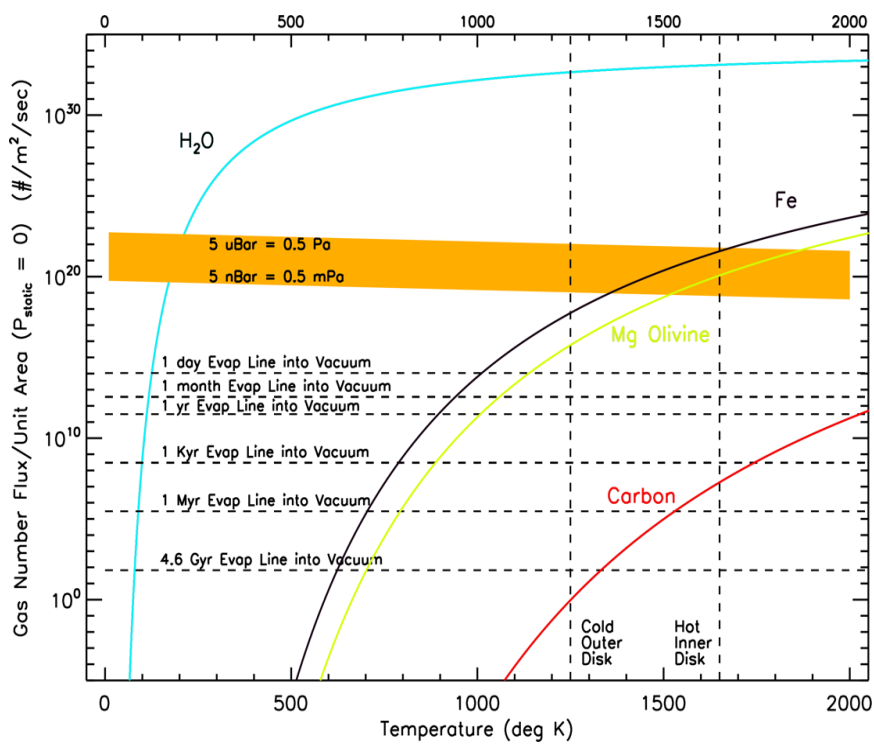

**Figure 16 – Sublimation loss rate and stability vs temperature curves of typical primitive solids found in proto-solar system material.** The dashed lines denote the levels of loss rate above which the solid sublimes and disappears into vacuum in a set time. Accretion disks around T Tauri stars like RW Aur A have a finite pressure, however, believed to be in the 0.5 mPa to 0.5 Pa range (Armitage+ 2001, Bouvier + 2013, Hartmann+ 2016; orange bar), and loss rates below this level correspond to external pressures where the solid is stable, while in the orange range the solid is in metastable equilibrium with the surrounding vapor. Loss rates above the orange bar cannot be resupplied by the surrounding gas, and the solid is again unstable. For the accretion disk temperature range of 1130 – 1650 K determined from the SpeX spectral observations, **water ice is never stable as a solid** at the expected pressures, and Fe is unstable in the innermost disk at ~1650K for all expected pressures. Solid rock (as proxied by Mg-olivine = Forsterite), is stable from 1130 to 1650 K at all but the innermost temperatures and lowest expected accretion disk pressures. Along with ultra-stable carbonaceous dust, olivine likely makes up much of the solid material in the disk. (Water ice sublimation data from Lisse *et al.* 2021 and references therein; Fe data from Alcock *et al.* 1984; C data from Darken & Gurry 1953; Olivine data from Costa *et al.* 2017).

The surrounding thermal environment in the innermost accretion disk at ~1650 K and 0.001 to 1 Pa then ensures that the Fe-rich core rubbilized material becomes gaseous Fe (Fig. 16), while the rocky mantle portion (if any) stays in the solid phase in the accretion disk. The Fe gas is then mobilized and thrown out into the system's jets that are rooted in the strong magnetic field lines threading through the inner edges of the accretion disk, or is guided by the same magnetic field lines into a shock at the top of the protostar's corona. The jet material plows into matter in the local environment around the excited CTTS (Figs. 1, 2b, 15) and also becomes shock heated and fluoresces as it travels away from the protostar at 100-200 km/sec, the Keplerian velocity range of the inner accretion disk wall.

While at first look the presence of abundant Fe and lack of rock forming elemental signatures would seem to indicate that an iron rich differentiated planetesimal core had just been catastrophically disrupted in 2014, the temperature and pressure conditions in the inner accretion disk are such that if a differentiated core + mantle object was disrupted, the rocky mantle portions would remain in (or quickly return to) the solid phase while the Fe-rich core portion would vaporize, leading to the same shocked Fe-rich vapor result. So while the progenitor planetesimal was differentiated, it was not necessarily also stripped of its mantle prior to the disruption event.

The temperature environment of the inner accretion disk also allows us to ignore the ~50% of the original condensed material (by mass; Desch 2015, McKinnon *et al.* 2017, Bierson *et al.* 2018) in any planetesimals condensed in icy fractions like $H_2O$, CO, $CO_2$, HCN, $CH_3OH$, etc. (Lisse *et al.* 2021), as this material would be sublimated into the accretion disk long before a planetesimal were to reach the 1200 – 1700 K temperatures (Fig. 16) found in the inner most sections of the RW Aur A accretion disk.

**5.3.2 Energetics:** An $\sim 10^{19}$ kg, Vesta-sized body, as suggested by the minimum mass of Fe II detected (Section 4.2), is large enough to have promoted differentiation of its interior into a mantle + core object via melting and density segregation (Kruijer *et al.* 2014 & references therein). On the other hand, the gravitational binding energy of a Vesta sized body is only ~0.07 MJ/kg (assuming the gravitational self-energy for a spherical body $U=(3/5)*GM^2/R_{body}$, $U/M \sim M/R \sim (4/3)\pi\rho R_{body}^2$, a value of U/M for the Earth of 37.5 MJ/kg and $R_{Vesta}/R_{Earth} \sim 1/24$). Thus disrupting such a body does not require vaporizing it, given that olivine and Fe take ~6 MJ/kg to vaporize. Since solid Fe is unstable in the inner portions of the accretion disk at r ~0.1 AU (Fig. 16), such a planetesimal had to form in more radially distant regions of the disk, r > 1 AU (Fig. 4), meaning that inspiralling migration is required to deliver the body to the inner accretion disk.

(Direct iron sublimation from an inwardly migrating "naked core" could potentially produce observable iron, but the sublimation rates from a solid iron body with radius of 100 - 1000 km, although appreciable at mm-cm/day (Fig. 16), would take 1000's of years to disperse the body and would not produce a large impulsive spike of new Fe. Very small (< 1 m radius) Fe particles vaporize on the order of years at the ~1650K temperatures we are finding for the innermost accretion disk, but would not be stable on longer time scales (Fig. 16), so could match the sudden uptick in system FeII - but they would have to be created in large numbers rapidly, e.g. via a collisional impact.)

**5.3.3 Dynamics:** The toy model provides a natural explanation for the observed jet outflow velocities and the relative ratio (~1:1) of new Fe mass emplaced into the system's jets vs. its stellar atmosphere. It also provides, by asking what rates of Fe II input will overwhelm the normal behavior of the system, an upper limit estimate for the duration of the FeII input event. The normal accretion rate of RW Aur A is ~ 5 x $10^{-8}$ M$_\odot$/yr (Valenti *et al.* 1993, Hartigan *et al.* 1995, White & Ghez 2001, Alencar *et al.* 2005, Facchini *et al.* 2016, Gárate *et al.* 2019, Koutoulaki *et al.* 2019), or ~3 x $10^{15}$ kg/s. Assuming a cosmogenic ratio of Fe/H ~ $10^{-4}$, this implies ~3 x $10^{11}$ kg/s of Fe accretion produces the very weak SpeX FeII lines of 2006-2007 (Fig. 13) and the Takami *et al.* 2020 pre-2014 weak FeII jet structures (Fig. 15). In order to exceed this Fe accretion rate and make the bright 2017 – 2020 FeII lines and structures, the accretion rate of the ~$10^{19}$ kg of new Fe must exceed 3 x $10^{11}$ kg/s, so it must have occurred in less than $10^{19}$ kg/3x$10^{11}$ kg/s = 3 x $10^7$ sec, or 1 year. More stringently, the brightest Nov 2018 SpeX FeII line reported here is ~10x brighter than the 2006/2007 lines, as is the Takami *et al.* (2020) post-outburst Feb 2017 FeII image versus the pre-outburst Dec 2014 image. This would imply the FeII was injected into the system's jets on the order of 0.1 year, or ~1 month. Timescales of ~1 month seem operative in the optical lightcurve trending of RW Aur A during the recent outburst (Rodriguez *et al.* 2013, 2016; Fig. 3).

**5.3.4 Implications.** If correct, the toy model has the strong implications that (A) newly created gas easily and quickly (on the order of weeks to months) merges into the accretion stream feeding into the protostar's atmosphere and the system's bipolar jets; (B) the outflow jets are rooted in the innermost accretion disk; (C) the jets are energized over their normal state by sudden gas excesses and overflows; and (D) the continued long-lived single peaked atmospheric emission FeII lines and low energy X-ray attenuation after ~5 years at the central protostar shows that the settling times in the protostar's outer atmospheric regions are long (and/or there is a continued steady, but not overwhelming, supply of fresh Fe gas from large, long-lived Fe-core boulders still accreting; c.f. Gárate *et al.* 2019). Thus the toy model predicts important impulse response function behavior for the system's jets and provides a useful framework future detailed time dependent modeling of its behavior.

**5.3.5 Caveats.** We caution that this model scenario suggests that the sensible Fe jet mass estimates (Section 4.2) are ***lower limits*** to the total Fe mass created by planetesimal destruction, with a large initial excess of Fe gas created by near-immediate vaporization of micron-sized iron grain rubble, leaving behind a remnant population of much larger meter+ sized boulders that slowly evaporate over years to decades. The amount of Fe-core mass left in these large, slowly evaporating boulders depends on the details of the catastrophic planetesimal breakup and the resulting rubble particle size distribution, but from analogous solar system and exodisk studies (e.g., Lisse *et al.* 2009, 2020), we estimate that the remnant large boulder mass will easily outmass the quickly sublimating fine particle rubble.

Similarly, we need to caution that the Fe measured with the spectroscopy presented in this paper is what is sensible along the Earth-RW Aur A line of sight in the jet outflow and in the protostar's atmosphere; we do not measure any initial Fe gas "spike" that may have already accreted into the protostar - so again the total Fe-excess mass amounts presented here ***are lower limits***, including the ~1:1 ratio of jet Fe-mass:total protostar atmosphere Fe-mass.

Finally, we present a speculative caveat concerning HeI. The partially bifurcated HeI 1.08 um line shown in Fig. 9 presents a bit of a puzzle. According to current models of CTTS structure (Hartmann *et al.* 2016 and references therein), hot HeI should reside in the protostar's outer atmospheric envelope, along with hot HI. At high galactic abundance vs. H (He: H ~ 0.1), this predicts strong single-peaked emission lines like the H Paschen and H Brackett features we detect (Figs. 7-9). Yet we also see a smaller bifurcated component to these lines that fades away over the course of 2018 – 2020 (Fig. 9), similar to the observed temporal variation of the Fe II lines, without any corresponding HI bifurcated component. This argues for important amounts of HeI in the jet outflow exhaust, but not important amounts of HI.

Given that we have concluded that the jet outflow signatures are coming from the disrupted and vaporized differentiated planetesimal core material, this would seem to imply that there was significant amounts of HeI in the disrupted planetesimal's core. This could not have been trapped HeI gas from the accretion disk of solar-like abundance, otherwise there should have been at least 10 times more HI gas present (ignoring the trapping efficiencies for H vs He in Fe-core material, which would favor proportionately more H; Zhang & Yin 2012, Roth *et al.* 2019, Bouhifd *et al.* 2020, Tagawa *et al.* 2021 and references therein). One possible explanation for its presence in the solid core material is that it is the byproduct of alpha particle decay over ~3 Myr from short lived radionuclides in the core material, although most of the important identified short lived radionuclides in the solar system's history, like $Al^{26}$ and $Fe^{60}$, decay via beta emission chains that do not create alpha particles and He. Another possibility is that excess primordial HI *is* present, along with excess primordial HeI, but that HI is launched into the jets very inefficiently. A focused future time domain study in HeI, HI, and FeII of RW Aur A's jets, to follow up on this mystery and determine if there is indeed an overabundance of HeI in them, seems in order.

## 6. Conclusions

Our NASA IRTF/SpeX spectral monitoring campaign has verified and extended a number of previous studies of the RW Aur A CTTS system. The use of long term medium resolution, large grasp, high precision spectral monitoring has allowed us to identify and determine 5 different spectral flux components in the data. In finding these 5 spectral components, the campaign has verified

1. The presence of ~4000K photospheric emission that remains constant in spectral slope and color throughout, but varies in amplitude, as previously noted by Facchini *et al.* 2016 & references therein; from this we can deduce that the cause of its variable flux amplitude must be the addition or subtraction of a neutral gray absorber/scatterer at 0. 7 – 1.3 μm wavelengths in the upper regions of the protostar's atmosphere, inside the inner edge of the accretion disk. The ~4000K emission color temperature we find, required to explain the stability of the observed flux from 0.7 to 1.3 μm, is similar to, but somewhat cooler, than the 4900K best-fit temperature found by Koutoulaki *et al.* 2019.

2. Associated single-peaked H-Paschen emission lines, that have amplitudes temporally correlated with the photospheric flux levels. These lines likely are formed also in the upper regions of the protostar's atmosphere, inside the inner edge of the accretion disk. Hot CO molecular gas sharing similar temperature also seems to be present at the top of the protostar's atmosphere; this gas extends into the inner parts of the accretion disk itself, forming a bridge between the two regions.

3. Highly variable (both in magnitude and in color temperature) thermal emission from the surrounding accretion disk at 1.7 – 3.3 µm. The magnitude of the accretion disk emission inversely correlates with its color temperature and with the photospheric and H-Paschen flux levels. The "puffed-up" accretion disk model proposed by Facchini *et al.* 2016 explains this very well – at times the accretion disk expands so much that its colder outer reaches geometrically block our line of sight view into the hot regions of the innermost system. Another possibility is that the outer edges of the accretion disk are not uniform in vertical extent and opacity, and what we see is modulated by rotation of thicker clouds through the LOS.

4. A stochastic appearance of strong, bifurcated Fe II, SiI, SI, SrI, and HeI emission lines from 2015 through early 2020, and their change into a smaller single peaked lines of ~1/3 the total EW by late 2020. The unique bifurcated nature of these lines links them to the accretion funnels and bipolar outflow jets of the system. From the measurements of the accretion disk at its lowest flux and highest color temperature epochs, we find the inner wall to be at ~1650 K, or 0.037 AU from a 1.6 $L_{solar}$ photospheric surface, while the co-rotation distance is a little larger at 0.042 AU. The Keplerian velocities at these two radii are 184 km/sec and 173 km/sec, respectively. Knots in the outflow jets have been measured moving with apparent velocity of 100 – 200 km/sec (Melnikov *et al.* 2009). The presence of Fe II and not FeI, and of HeI, indicate the emitting material has gone through a strong shock (Koo *et al.* 2016). Gas dominated by Fe, Si, S, and C atoms is expected as the product of vaporization of a planetesimal core with abundances like the Earth's, is consistent with the proposed model of Günther *et al.* (2018). The lack of any associated Al, Ca, Mg, or Na emission lines rules out the presence of much planetesimal mantle or crustal material.

5. Hot CO gas emission coincident with regions right inside the accretion disk inner wall and is clearly seen from 2006 through 2019, but it disappears by late 2020 when the system appears to be returning back to normal. The temporal signature for the CO emission is different from any of the other flux sources, being relatively steady in amplitude until disappearing totally in late 2020, suggesting that it is sourced differently than HI in the same region. Its disappearance within the space of a year suggests that hot CO's residence time inside the accretion disk is short.

6. All of these phenomenological spectral findings demonstrate a highly excited and driven CTTS system at ~3 Myr, very unlike the spectral behavior seen for the "relaxed", slowly developing RW Aur B close T Tauri binary companion. RW Aur A is acting like a "re-born", few 0.1 Myr T Tauri star with strong jets, emission lines, and obscuring dusty envelope, while RW Aur B is acting its age and transitioning to a WTTS.

7. A physical model motivated by the observations of abundant new Fe in RW Aur A since 2014 imply that differentiated, migrating planetesimals with radius > 100 km undergoing catastrophic disruption in the hottest, innermost regions of the accretion disk exist in the system. The Fe-rich core portions of the disrupted planetesimal vaporize, creating a sudden overabundance of Fe. This overpressure of iron is then blown out into the system's bipolar "exhaust" jets and dumped into the protostar's extended envelope and corona, where it obscures the normal low-energy accretion-driven X-rays while making new 6-7 keV Fe-X-ray emission. A 3 Myr age for RW Aur A (Dodin *et al.* 2020) is consistent with the presence of differentiated bodies in our solar system: the isotopic measurements of Kruijer *et al.* 2014 indicate that the differentiation of iron meteorite parent bodies was protracted and occurred from ~0.7-1.3 Myr after the first solar system solids formed, and planetesimals accreting in the first few Myr larger than 40 km in diameter were likely differentiated (Hevey & Sanders 2006).

## 7. Acknowledgements

This paper was based on observations taken with the NASA IRTF/SpeX 0.8–5.5 Micron Medium-Resolution Spectrograph and Imager, funded by the National Science Foundation and NASA and operated by the NASA Infrared Telescope Facility. C.M. Lisse gratefully acknowledges support for this work from the NASA Chandra and NASA NExSS programs.